\documentclass[useAMS,usenatbib]{mn2e}
\pdfoutput=1
\usepackage{aas_macros}
\usepackage{epsfig}
\usepackage{amsmath}
\usepackage{amssymb}
\usepackage{comment}
\usepackage{caption}
\usepackage{leftidx}
\usepackage{natbib}
\usepackage[rgb]{xcolor}
\definecolor{MyGreen}{rgb}{0.0,0.6,0.3}
\definecolor{MyPurple}{rgb}{0.6,0,0.3}
\usepackage{fancyref}
\usepackage{hyperref}
\hypersetup{colorlinks=true,citecolor=MyGreen,linkcolor=MyPurple,urlcolor=blue}

\def\beq{\begin{equation}}
\def\eeq{\end{equation}}
\def\ba{\begin{eqnarray}}
\def\ea{\end{eqnarray}}
\def\bal{\begin{align}}
\def\eal{\end{align}}
\def\bxi{{\mbox{\boldmath $\xi$}}}
\def\bnab{{\mbox{\boldmath $\nabla$}}}

\begin{document}

\title[Magnetic Structure] {Linking the Interiors and Surfaces of Magnetic Stars}

\author[Fuller \& Mathis]{
Jim Fuller$^{1}$\thanks{Email: jfuller@caltech.edu} and St\'ephane Mathis$^{2}$
\\$^1$TAPIR, Mailcode 350-17, California Institute of Technology, Pasadena, CA 91125, USA
\\$^2$Université Paris-Saclay, Université Paris Cité, CEA, CNRS, AIM, F-91191, Gif-sur-Yvette, France
}

\label{firstpage}
\maketitle

\begin{abstract}

Strong magnetic fields are observed in a substantial fraction of upper main sequence stars and white dwarfs. Many such stars are observed to exhibit photometric modulations as the magnetic poles rotate in and out of view, which could be a consequence of magnetic perturbations to the star's thermal structure. The magnetic pressure is typically larger than the gas pressure at the star's photosphere, but much smaller than the gas pressure in the star's interior, so the expected surface flux perturbations are not clear. We compute magnetically perturbed stellar structures of young $3 \, M_\odot$ stars that are in both hydrostatic and thermal equilibrium, and which contain both poloidal and toroidal components of a dipolar magnetic field as expected for stable fossil fields. This provides semi-analytical models of such fields in baroclinic stably stratified regions. The star's internal pressure, temperature, and flux perturbations can have a range of magnitudes, though we argue the most likely configurations exhibit flux perturbations much smaller than the ratio of surface magnetic pressure to surface gas pressure, but much larger than the ratio of surface magnetic pressure to central gas pressure. The magnetic pole is hotter than the equator in our models, but a cooler magnetic pole is possible depending on the magnetic field configuration. The expected flux variations for observed field strengths are $\delta L/L \! \lesssim \! 10^{-6}$, much smaller than those observed in magnetic stars, suggesting that observed perturbations stem from changes to the emergent spectrum rather than changes to the bolometric flux.

\end{abstract}


\section{Introduction}

Magnetic fields produce a substantial impact on the appearance of a star. In stars with convective envelopes, star spots are well known as regions of high magnetic field strength and low temperature. Some stars with radiative envelopes are also known to host strong magnetic fields (e.g., \citealt{Morel:2014,Wade:2016,Shultz:2019}), and such stars frequently exhibit photometric variability at their rotation periods, suggesting that their emergent flux is altered by the magnetic fields. However, it is not clear how the magnetic fields actually affect the stellar structure and emergent flux, or whether photometric modulations can be used to learn about the strength of the star's internal magnetic fields. 

One might expect magnetic fields to affect the photospheric temperature if the magnetic pressure is comparable to the photospheric gas pressure. Indeed, \cite{Cantiello:11} argued that magnetic spots on massive stars should be hot because they would have lower gas pressures (and therefore lower densities), allowing us to see deeper into the star where the temperature is higher. Relatively modest magnetic fields of $B \gtrsim 100 \, {\rm G}$ are required for large flux perturbations in this scenario. Observed photometric modulations from stars with stronger magnetic fields, such as Ap stars that exhibit photometric modulation with $\sim$1-3\% amplitudes \citep{hummerich:18}, are much smaller than naively predicted from this scenario.

The reason is that a star's radiative flux will change in response to the magnetic perturbation (i.e., the magnetic hot spot will cool off) until the star finds a new radiative equilibrium. Long-lived magnetic fields will thus produce 
much different effects than transient fields arising from magnetic activity. To compute the perturbed structure of a star with a stable magnetic field, we must find a structure that is in both hydrostatic equilibrium and radiative equilibrium. For rotating stars, it is well known that no state of radiative equilibrium exists for solid body rotation, so that stars must either have very special rotation profiles \citep{vonZeipel:1924,Busse:1981,Rieutord2006}, or have currents that advect heat \citep{eddington:29,sweet:50} and restore a state of equilibrium (see \citealt{maeder:99}, \citealt{decressin:2009}, \citealt{mathis:2013} for useful synopsis). Our goal in this paper is to compute the special magnetic field profiles that allow for radiative equilibrium without requiring currents within the star.

In addition, the magnetic field configuration must be a \textit{stable} equilibrium that does not unravel via magnetic instabilities such as those of purely toroidal fields \citep{tayler:73} and purely poloidal fields \citep{markeytayler:73}. Several works \citep{Braithwaite2004,Braithwaite_2006,Braithwaite2008,broderick:08,lyutikov:10,duez:10,duez:10b,akgun:13,becerra:22b} have computed stable magnetic equilibria through analytic calculations or numerical simulations. These works agree that purely poloidal or toroidal magnetic field configurations are unstable, and that stable equilibria require similar toroidal and poloidal field strengths \citep{tayler:80,braithwaite:09}. Stable stratification (i.e., non-barotropic stars) is also required for long-term stability \citep{lander:12,akgun:13,becerra:22}. Magnetic configurations decrease their magnetic energy but approximately conserve their magnetic helicity as they form and evolve \citep{Braithwaite2008}. This is the so-called selective decay as observed in plasmas in the laboratory \citep{taylor:1974}. Hence, recent works have computed stable field configurations through variational techniques that minimize total energy while conserving magnetic helicity \citep{broderick:08,duez:10}. 

However, all of these works have assumed barotropic perturbations such that the perturbed temperature is directly proportional to the perturbed pressure. While these configurations are in hydrostratic equilibrium, they are not typically in thermal equlibrium, meaning that heat will be transported from high temperature to low temperature regions, changing the gas pressure and therefore the magnetically perturbed stellar structure. This heat transport will occur on a thermal time (typically $\sim \! 10^6 \, {\rm yr}$ in A-type stars), much shorter than Ohmic diffusion time scales ($\sim \! 10^{10} \, {\rm yr}$) and main sequence life times ($\sim \! \! 10^9 \, {\rm yr}$). Hence, real stars will approach a state close to hydrostatic equilibrium and thermal equilibrium, which has not been considered in recent literature. Accounting for thermal equilibrium is crucial for predicting the long-term equilibria of stars \citep{Reisenegger:2009}, and observational manifestations such as the perturbed surface flux.

Calculations of equilibrium magnetic configurations date back to the 1950s (e.g., \citealt{chandrasekhar:56,chandrasekhar:56b}). Subsequent work included the effects of centrifugal distortion and meridional flows (e.g., \citealt{ostriker:68,mestel:77}).
Interestingly, works dating back to the 1960s \citep{monaghan:66,davies:68,wright:69,moss:1973,moss:79,li:06} have attempted to compute the structures of stars in both hydrostatic and thermal equilibrium. However, much of this work is either difficult to interpret, does not discuss the perturbed thermal structure and surface flux, does not consider fields with both poloidal and toroidal components, or has simply been forgotten. The goal of this paper is to provide updated calculations of magnetically distorted stars in stable hydrostatic and thermal equilibrium for realistic stellar structures, and to discuss the observational implications. 

In this paper we primarily focus on application to upper main sequence stars of $M \! \gtrsim \! 1.5 \, M_\odot$ with convective cores and radiative envelopes. Much of the physics studied here could also apply to radiative stars such as white dwarfs and the radiative cores of red giants where internal fields can be detected through asteroseismology \citep{Garcia2014,stello:16,Li:2022} because of their impact on stellar oscillations \citep{Fulleretal2015,Lecoanetetal2017,Loi:2021,Bugnet:2021,Mathis:2021}. Magnetic upper main sequence stars have typical surface field strengths of $\sim$1 kG (see \citealt{braithwaite:17} for a review) and surface magnetic pressures larger than surface gas pressures, such that magnetic forces could be strong. The surface magnetic morphologies are observed to be diverse: they can be complex or simple, axisymmetric or non-axisymmetric, poloidal or toroidal \citep[][]{LandstreetMathys2000,Kochukhovetal2011,Shultz:2019}. However, as a first step, in this work we examine simple dipolar magnetic configurations that produce quadrupolar temperature/pressure perturbations, which likely dominate observable photometric modulations.

\section{Equilibrium Structure}
\label{sec:eq}

Our goal is to calculate the structure of a magnetized star in hydrostatic and thermal equilibrium, considering non-force-free magnetic fields. We begin from the equation for magnetohydostratic equilibrium
\beq
\label{eq:hydro}
- \bnab P - \rho \bnab \Phi + \frac{ (\bnab \times {\bf B}) \times {\bf B}}{4 \pi} = 0 \, ,
\eeq
where $P$ is the pressure, $\rho$ the density, $\Phi$ the gravitational potential, and ${\bf B}$ the magnetic field. We choose to work in standard spherical coordinates with $r$ the radius and $\theta$ the colatitude. Following \cite{duez:10}, we decompose the magnetic field into a poloidal magnetic stream function $\Psi$ and a toroidal magnetic flux $F$ via
\beq
{\bf B} = \frac{ \sqrt{4 \pi} }{r \sin \theta} \bnab \Psi \times \hat{\phi} + \frac{\sqrt{4 \pi}}{r \sin \theta} F \hat{\phi} \, .
\eeq
This means that
\beq
\label{eq:br}
B_r = \frac{ \sqrt{4 \pi} }{r^2 \sin \theta} \frac{\partial \Psi}{\partial \theta} \, ,
\eeq
\beq
\label{eq:bt}
B_\theta = \frac{ -\sqrt{4 \pi} }{r \sin \theta} \frac{\partial \Psi}{\partial r} \, ,
\eeq
\beq
\label{eq:bp}
B_\phi = \frac{ \sqrt{4 \pi}}{r \sin \theta} F \, .
\eeq
We consider an axisymmetric magnetic field such that $\Psi$ and $F$ are independent of $\phi$. It can easily be verified that this field always satisfies $\bnab \cdot \mathbf{B}=0$.

The $\phi$-component of equation \ref{eq:hydro} becomes 
\beq
\label{eq:hydrophi}
- \frac{\partial F}{\partial \theta} \frac{\partial \Psi}{\partial r} + \frac{\partial F}{\partial r} \frac{\partial \Psi}{\partial \theta } = 0 \, .
\eeq
This is always satisfied if $F$ is a function of $\Psi$, or in other words, the poloidal flux $\Psi$ uniquely determines $F$ and hence the toroidal field. \cite{broderick:08} and \cite{duez:10} show that the lowest energy state has
\begin{equation}
\label{eq:fpsi}
F = \frac{\lambda}{R} \Psi \, ,
\end{equation}
where the constant $\lambda$ determines the magnetic helicity, and $R$ is the stellar radius. This clearly satisfies equation \ref{eq:hydrophi}. This also results if we assume that $F$ and $\Psi$ have the same angular form, such that equation \ref{eq:hydrophi} reduces to 
\begin{equation}
    \frac{1}{F} \frac{d F}{dr} = \frac{1}{\Psi} \frac{d \Psi}{dr} \, .
\end{equation}
The solution to this equation is equation \ref{eq:fpsi}, for a constant $\lambda$ that is independent of radius.

Writing out the hydrostatic equilibrium condition and using equation \ref{eq:fpsi}, the radial component of equation \ref{eq:hydro} becomes
\begin{equation}
\label{eq:hydror}
- \frac{ 1 }{r^2 \sin^2 \theta} \bigg[ \frac{\lambda^2}{R^2} \Psi + \Delta^* \Psi \bigg] \frac{\partial \Psi}{\partial r} = \frac{\partial P}{\partial r} + \rho \frac{\partial \Phi}{\partial r} \, ,
\end{equation}
while the $\theta$-component is
\begin{equation}
\label{eq:hydrotheta}
- \frac{ 1 }{r^2 \sin^2 \theta} \bigg[ \frac{\lambda^2}{R^2} \Psi + \Delta^* \Psi \bigg] \frac{\partial \Psi}{\partial \theta} = \frac{\partial P}{\partial \theta} + \rho \frac{\partial \Phi}{\partial \theta} \, .
\end{equation}
Here, $\Delta^*$ is the ``Grad-Shafranov" or ``five-dimensional Laplacian" operator, defined as
\begin{align}
    \Delta^* &= \frac{\partial^2}{\partial r^2} + \frac{\sin \theta}{r^2} \frac{\partial}{\partial \theta} \bigg(\frac{ 1 }{\sin \theta} \frac{\partial}{\partial \theta} \bigg) \nonumber \\
    &= \frac{\partial^2}{\partial r^2} + \frac{1-\mu^2}{r^2} \frac{\partial^2}{\partial \mu^2} 
    \, ,
\end{align}
where $\mu = \cos \theta$.

It is immediately evident from equations \ref{eq:hydror} and \ref{eq:hydrotheta} that the field is force-free if
\begin{equation}
\label{eq:forcefree}
    \frac{\lambda^2}{R^2} \Psi + \Delta^* \Psi = 0 \, .
\end{equation}
This results in a simple linear eigenvalue calculation for the magnetic potential $\Psi$, and is the type of field considered by \cite{broderick:08}. The work of \cite{duez:10} considers a non-force-free field such that 
\begin{equation}
\label{eq:duez}
    \frac{\lambda^2}{R^2} \Psi + \Delta^* \Psi = - \beta \rho r \sin^2 \theta \, ,
\end{equation}
where $\beta$ is a constant that determines the strength of the magnetic force. As an example, the case with $\lambda=0$ and $\beta=0$ corresponds to a force-free poloidal dipole field. This can be seen from equation \ref{eq:forcefree}, whose solution has $\Psi \propto r^{-1}$ and hence $B \propto r^{-3}$ in that case. A non-zero value of $\beta$ alters both the magnetic forces and the radial profile of the field.

Substitution of equation \ref{eq:duez} into equations \ref{eq:hydror} and \ref{eq:hydrotheta} yields
\begin{equation}
\label{eq:beta}
    \beta \rho \bnab \Psi = \bnab P + \rho \bnab \Phi \, .
\end{equation}
Hence there is a direct relationship between the magnetic flux and the pressure perturbation for barotropic perturbations. Remarkably, the non-linear equations \ref{eq:hydror} and \ref{eq:hydrotheta} have been transformed into a linear relationship between $\Psi$ and $P$. Taking the curl of equation \ref{eq:beta} yields
\begin{equation}
\label{eq:baro}
    \bnab \rho \times \bnab P = 0 \, .
\end{equation}
This implies that $P$ is a function of $\rho$, and hence equation \ref{eq:duez} is a solution for a barotropic equation of state such that density and pressure perturbations are directly proportional to each other.

In our work, we want to consider non-force-free fields that produce non-barotropic density and pressure perturbations. Hence, we cannot use the approximations made in \cite{broderick:08} or \cite{duez:10}, and we shall see that this generally leads to a series of non-linear differential equations that relate the magnetic field to density, temperature, and pressure perturbations.
In our calculations, we parameterize the strength of the magnetic forces via a parameter $\beta$. Assuming barotropic perturbations entails that $\beta$ (as defined in equation \ref{eq:duez}) is a constant within the star. Accounting for radiative diffusion, this is no longer the case. Nonetheless, we shall see below that we still require a parameter to specify the strength of the magnetic forces, which in practice is determined by the boundary conditions. Since our stellar model has a convective core where equation \ref{eq:duez} is a good approximation, we label our structures based on the resulting $\beta$ in the convective core.

\subsection{Electric Currents}
\label{sec:j}

The current density is
\begin{align}
\label{eq:j}
    {\bf j} &= \frac{\bnab \times {\bf B}}{4 \pi} \nonumber \\ 
    &= \frac{\lambda}{4 \pi R} {\bf B_r} \hat{r} + \frac{\lambda}{4 \pi R} {\bf B_\theta} \hat{\theta} - \frac{1}{\sqrt{4 \pi} r \sin \theta} \Delta^* \Psi \hat{\phi} \, .
\end{align}
The force-free field of equation \ref{eq:forcefree} occurs when ${\bf j}$ is parallel to ${\bf B}$. Force-free fields are also obtained when $\Psi=0$ (no magnetic field) or from current-free fields, which only occur when both $\lambda=0$ and and $\Delta^* \Psi=0$. Note that the radial current is proportional to the radial magnetic field, hence a vanishing radial current near the surface of the star requires a vanishing radial field $B_r$, which in turn requires $\Psi$ to vanish at the stellar surface.

\subsection{Hydrostatic Equilibrium}

In order to determine an equilibrium state, we use a linear approximation such that perturbations to background quantities ($\rho$, $P$, etc.) are considered to be small. The linearized version of the radial momentum equation (\ref{eq:hydror}) is 
\begin{equation}
\label{eq:hydror2}
- \frac{ 1 }{r^2 \sin^2 \theta} \bigg[ \frac{\lambda^2}{R^2} \Psi + \Delta^* \Psi \bigg] \frac{\partial \Psi}{\partial r} = \frac{\partial}{\partial r} \delta P + \rho \frac{\partial}{\partial r} \delta \Phi + g \delta \rho \, ,
\end{equation}
while the $\theta$-component of equation (\ref{eq:hydrotheta}) is
\begin{equation}
\label{eq:hydrotheta2}
- \frac{ 1 }{r^2 \sin^2 \theta} \bigg[ \frac{\lambda^2}{R^2} \Psi + \Delta^* \Psi \bigg] \frac{\partial \Psi}{\partial \theta} = \frac{\partial}{\partial \theta} \delta P + \rho \frac{\partial }{\partial \theta} \delta \Phi \, .
\end{equation}
Here, $\delta$ indicates an Eulerian perturbation, and we have used a background in hydrostatic equilibrium with $dP/dr = - \rho g$, and $g = d \Phi/dr$.
%In real stars, the magnetic pressure is comparable to the gas pressure only very near the stellar surface, so we only expect large density changes in the outer layers which have very little mass. Hence, we adopt the Cowling approximation in which $\delta \Phi$ can be ignored, eliminating the last term in equations \ref{eq:hydror2} and \ref{eq:hydrotheta2}.

\begin{figure}
\includegraphics[scale=0.3]{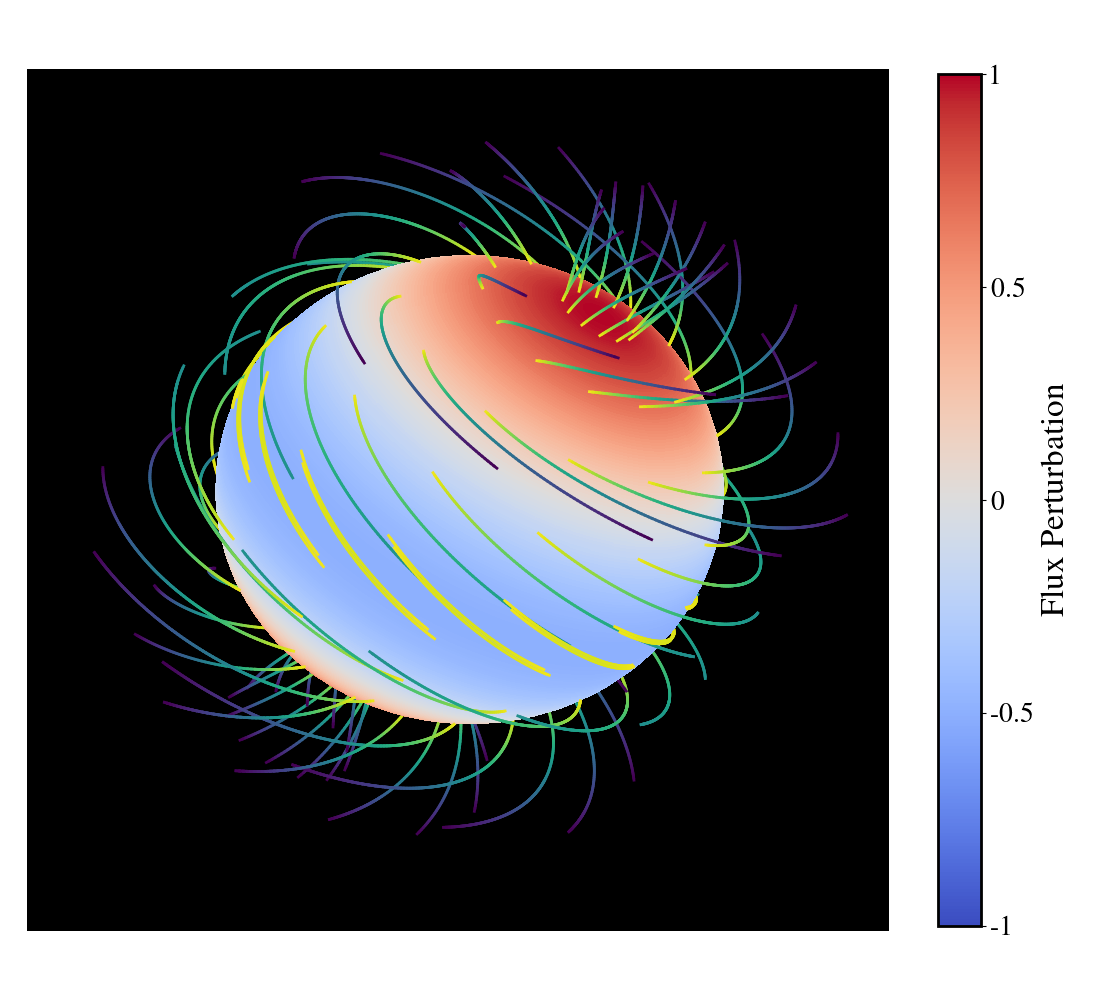}
\caption{\label{fig:magstruc} Magnetic field lines of a star for a toroidal field comparable to the poloidal field ($\lambda^2=10)$. Magnetic field lines are colored by field strength, while the color shading indicates the radiative flux perturbation at a given radius.
}
\end{figure}

We next turn to the angular dependence of the magnetic field and the perturbations to the stellar structure. Decomposing $\Psi$ into eigenvalues of the horizontal component of $\Delta^*$ requires
\begin{equation}
\label{eq:g}
(1 - \mu^2) \frac{\partial^2}{\partial \mu^2} g_\ell(\mu) = - \ell (\ell +1) g_\ell(\mu) \, ,
\end{equation}
where $g_\ell$ is the angular eigenfunction corresponding to the eigenvalue $-\ell(\ell+1)$ of the operator on the left hand side of equation \ref{eq:g}. The full response is $\Psi = \sum_\ell \Psi_\ell g_\ell(\mu)$. As discussed in \cite{duez:10}, the eigenvalue equation above has solutions 
\begin{equation}
    g_\ell(\mu) = (1-\mu^2) P_{\ell-1}(\mu) \, ,
\end{equation}
where $P_\ell$ is a Legendre polynomial.

For the lowest order (dipole) solution with $\ell=1$, we have $g_\ell(\mu) = (1-\mu^2)$ and hence $\Psi = \Psi_\ell(r) \sin^2 \theta$. Plugging this into equation \ref{eq:hydror2} yields 
\begin{align}
\label{eq:hydror3}
&-\frac{ \sin^2(\theta) }{r^2} \bigg[ \frac{\partial^2}{\partial r^2} \Psi_\ell(r) - \frac{\ell(\ell+1)}{r^2} \Psi_\ell(r) + \frac{\lambda^2}{R^2} \Psi_\ell(r) \bigg] \frac{\partial \Psi_\ell(r)}{\partial r} \nonumber \\
 &= \frac{\partial}{\partial r} \delta P + \rho \frac{\partial}{\partial r} \delta \Phi + g \delta \rho \,.
\end{align}
We see that the perturbed density, pressure, and potential must have angular form $\delta P \propto \sin^2 (\theta)$, which is a combination of the $\ell=2$ and $\ell=0$ spherical harmonics. Hence a dipole magnetic field induces both radial and quadrupole components to the star's distortion. \autoref{fig:magstruc} illustrates the geometry of a dipolar magnetic field with helicity $\lambda^2=10$ that induces quadrupolar flux perturbations. The quadrupole ($\ell=2$) component of the field induces $\ell=4$, $\ell=2$, and $\ell=0$ components of the stellar distortion. 

We thus have the unfortunate situation that the angular eigenfunctions of the magnetic and hydrodynamic variables are not the same, meaning that the radial and angular parts of the response cannot generally be separated. In this work, we limit ourselves to dipole magnetic field configurations, which induce $\ell=0$ and $\ell=2$ components to the stellar structure perturbations. We are not interested in the $\ell=0$ component of the stellar response, as it is the non-radial magnetic distortions that draw our focus. Hence, from here forward, we consider a dipole ($\ell=1$) magnetic field and the quadrupolar ($\ell=2$) component of the stellar response.

Letting the pressure response be 
\begin{equation}
\label{eq:decomp}
\delta P = a_0 \delta p_0(r) Y_{0 0}(\theta) + a_2 \delta p_2(r) Y_{2 0}(\theta)
\end{equation}
and setting the angular dependence equal to $\sin^2 \theta$ requires $a_2 = -\sqrt{16 \pi/45}$. The radial component of the response is then $a_0 \delta p_0(r) = (4 \sqrt \pi/3) \delta p_2(r)$.
Dropping the $(r)$ dependence and subscripts of $\Psi_\ell$ and $\delta p_2$ for simplicity, equations \ref{eq:hydror2} and \ref{eq:hydrotheta2} can be written 
\begin{align}
\label{eq:hydror4}
&\bigg[ \frac{\partial^2}{\partial r^2} \Psi - \frac{\ell(\ell+1)}{r^2} \Psi + \frac{\lambda^2}{R^2} \Psi \bigg] \frac{\partial \Psi}{\partial r} \nonumber \\
&= - r^2 \frac{\partial \delta p}{\partial r} - r^2 \rho \frac{\partial}{\partial r} \delta \Phi - r^2 g \delta \rho \, ,
\end{align}
\begin{equation}
\label{eq:hydrotheta4}
\bigg[ \frac{\partial^2}{\partial r^2} \Psi - \frac{\ell(\ell+1)}{r^2} \Psi + \frac{\lambda^2}{R^2} \Psi \bigg]  \Psi = - r^2 \delta p - r^2 \rho \delta \Phi  \, ,
\end{equation}
and it is now understood that these equations are only valid for $\ell=1$ and $\delta p$, $\delta \Phi$ etc. refer to the quadrupolar part of the stellar response. Equation \ref{eq:hydrotheta4} can be substituted into equation \ref{eq:hydror4} to obtain
\begin{equation}
\label{eq:hydror5}
\big( \delta p + \rho \delta \Phi \big) \frac{\partial \Psi}{\partial r} = \bigg[\frac{\partial \delta p}{\partial r} + \rho \frac{\partial}{\partial r} \delta \Phi + g \delta \rho \bigg] \Psi \, .
\end{equation}
However, we have divided by $\Psi$ to obtain this equation, so we must be wary of solutions that cross $\Psi=0$.

The gravitational potential perturbation is given by Poisson's equation,
\begin{equation}
    \nabla^2 \delta \Phi = 4 \pi G \delta \rho \, .
\end{equation}
This can be written in terms of two first-order equations,
\begin{equation}
\label{eq:pos1}
    \frac{\partial}{\partial r} \delta \Phi - \delta \Phi' = 0 \, ,
\end{equation}
\begin{equation}
\label{eq:pos2}
    \frac{\partial}{\partial r} \delta \Phi' + \frac{2}{r} \delta \Phi' - \frac{(\ell+1)(\ell+2)}{r^2} \delta \Phi - 4 \pi G \delta \rho = 0 \, .
\end{equation}

\subsection{Thermal Equilibrium}

We next turn to the equations of thermal equilibrium. Energy conservation requires 
\beq
\label{eq:dsdt}
\rho T \frac{d s}{dt} = \rho \epsilon- \bnab \cdot {\bf F} \,, 
\eeq
where $T$ is temperature, $s$ is specific entropy, $\epsilon$ is the specific energy generation rate, and ${\bf F}$ is the energy flux. In thermal equilibrium, the entropy is constant, and the background state only has a radial flux $4 \pi r^2 F = L$, which entails that $dL/dr = 4 \pi \rho r^2 \epsilon$. Additionally, the energy flux is
\begin{equation}
\label{eq:flux}
{\bf F} = - \chi \bnab T
\end{equation}
where
\begin{equation}
\label{eq:chi}
    \chi = \frac{4 a c T^3}{3 \kappa \rho} \, .
\end{equation}
is the thermal diffusivity. This means that the background temperature gradient is $dT/dr = - L/(4 \pi r^2 \chi)$. 

For a perturbation in thermal equilibrium, the Eulerian perturbation of equation \ref{eq:dsdt} is 
\begin{equation}
\label{eq:divdf}
    \bnab \cdot \delta {\bf F} - \delta \rho \epsilon - \rho \delta \epsilon = 0 \, .
\end{equation}
The Eulerian perturbation of equation \ref{eq:flux} is 
\begin{equation}
\label{eq:df}
    {\bf \delta F} =  - \delta \chi \frac{dT}{dr} {\hat {\bf r}} - \chi \bnab \delta T \, .
\end{equation}
Taking the horizontal divergence of the horizontal part of this equation yields
\begin{equation}
    \bnab_\perp \cdot {\bf \delta F}_\perp =  - \chi \nabla_\perp^2 \delta T \, ,
\end{equation}
where $\nabla_\perp^2 = -(\ell+1)(\ell+2)/r^2$ since we are considering perturbations with spherical harmonics of degree $\ell+1$. Plugging this into equation \ref{eq:divdf} and using $4 \pi r^2 \delta F_r = \delta L$, we obtain
\begin{equation}
    \frac{\partial}{\partial r} \delta L + 4 \pi (\ell+1)(\ell+2) \chi \delta T - 4 \pi \rho r^2 \bigg( \frac{\delta \rho}{\rho} \epsilon + \delta \epsilon \bigg)  = 0 \, .
\end{equation}
This can be rewritten
\begin{equation}
    r \frac{\partial}{\partial r} \frac{\delta L}{L_s} + \frac{4 \pi (\ell+1)(\ell+2) \chi T r}{L_s} \frac{\delta T}{T} - \frac{r}{L_s}\frac{dL}{dr} \bigg( \frac{\delta \rho}{\rho} + \frac{\delta \epsilon}{\epsilon} \bigg)  = 0 \, ,
\end{equation}
where $L_s$ is the star's surface luminosity. The energy generation perturbation can be expanded as $\delta \epsilon/\epsilon = \epsilon_T \delta T/T + \epsilon_\rho \delta \rho/\rho$, where $\epsilon_T = (\partial \ln \epsilon/\partial \ln T)_\rho$ and  $\epsilon_\rho = (\partial \ln \epsilon/\partial \ln \rho)_T$, yielding
\begin{align}
    &r \frac{\partial}{\partial r} \frac{\delta L}{L_s} + \frac{4 \pi (\ell+1)(\ell+2) \chi T r}{L_s} \frac{\delta T}{T} \nonumber \\ &- \frac{r}{L_s}\frac{dL}{dr} \bigg( (1 + \epsilon_\rho) \frac{\delta \rho}{\rho} + \epsilon_T \frac{\delta T}{T} \bigg)  = 0 \, .
\end{align}

The radial component of equation \ref{eq:df} can be written
\begin{equation}
    \frac{\delta F_r}{F} = \bigg(3 \frac{\delta T}{T} - \frac{\delta \kappa}{\kappa} - \frac{\delta \rho}{\rho} \bigg) - \frac{\chi}{F} \frac{\partial \delta T}{\partial r} \, .
\end{equation}
Using $\delta \kappa/\kappa = \kappa_T \delta T/T + \kappa_\rho \delta \rho/\rho$, where $\kappa_T = (\partial \ln \kappa/\partial \ln T)_\rho$ and  $\kappa_\rho = (\partial \ln \kappa/\partial \ln \rho)_T$, this can be written as 
\begin{equation}
    r \frac{\partial}{\partial r} \bigg(\frac{\delta T}{T}\bigg) + \frac{L}{4 \pi r \chi T} \bigg[ \frac{\delta L}{L} - (4 - \kappa_T) \frac{\partial T}{T} + (1 + \kappa_\rho) \frac{\delta \rho}{\rho} \bigg] = 0.
\end{equation}

Finally, we require an equation of state to close the system of equations. This is given by
\begin{equation}
\label{eq:eos}
    \frac{\delta P}{P} - \chi_T \frac{\delta T}{T} - \chi_\rho \frac{\delta \rho}{\rho} = 0 \, ,
\end{equation}
where $\chi_T = (\partial \ln P/\partial \ln T)_\rho$ and  $\chi_\rho = (\partial \ln P/\partial \ln \rho)_T$ and are determined by the equation of state. This is valid if the composition is uniform, otherwise there will be an additional term in equation \ref{eq:eos}, which we discuss in Section \ref{sec:discussion}.

%\subsection{Derivation of Equilibrium Equations}
%\label{sec:rad}

\subsection{Equations and Boundary Conditions}
\label{sec:equations}

Putting everything together, we have a system of equations that can be solved for six variables: $\Psi$ and its radial derivative $\Psi'=\partial \Psi/\partial r$, and the Eulerian perturbations $\delta P$, $\delta \rho$, $\delta T$, and $\delta L$. The equations can be written
\beq
\label{eq:dpsidr}
\frac{\partial}{\partial r} \Psi - \Psi' = 0 \, ,
\eeq
\beq
\label{eq:dpsipdr}
r^2 \Psi \frac{\partial}{\partial r} \Psi' + \bigg(\frac{\lambda^2 r^2}{R^2} - \ell(\ell+1) \bigg) \Psi^2 + r^4 P \frac{\delta P}{P} + r^4 \rho \delta \Phi = 0 \, ,
\eeq
\begin{align}
\label{eq:dpdr}
& r \Psi \frac{\partial}{\partial r} \frac{\delta P}{P} + \frac{\rho g r}{P} \Psi \frac{\delta P}{P} + \frac{r \rho}{P} \Psi \frac{\partial}{\partial r} \delta \Phi + \frac{\rho g r}{P} \Psi \frac{\delta \rho}{\rho} \nonumber \\  &- r \Psi' \bigg(\frac{\delta P}{P} + \frac{\rho \delta \Phi}{P} \bigg)  = 0 \, ,
\end{align}
\begin{equation}
\label{eq:dtdr}
    r \frac{\partial}{\partial r} \bigg(\frac{\delta T}{T}\bigg) + \frac{L}{4 \pi r \chi T} \bigg[ \frac{\delta L}{L} - (4 - \kappa_T) \frac{\delta T}{T} + (1 + \kappa_\rho) \frac{\delta \rho}{\rho} \bigg] = 0 \, ,
\end{equation}
\begin{align}
\label{eq:dldr}
    &r \frac{\partial}{\partial r} \frac{\delta L}{L_s} + \frac{4 \pi (\ell+1)(\ell+2) \chi T r}{L_s} \frac{\delta T}{T} \nonumber \\ &- \frac{r}{L_s}\frac{dL}{dr} \bigg( (1 + \epsilon_\rho) \frac{\delta \rho}{\rho} + \epsilon_T \frac{\delta T}{T} \bigg)  = 0 \, ,
\end{align}
along with Poisson's equation (equations \ref{eq:pos1} and \ref{eq:pos2}) and the equation of state (equation \ref{eq:eos}). This system contains seven first-order differential equations, two of which are non-linear. They depend on the field geometry $\ell$ (assumed to be $\ell=1$ here), and the magnetic helicity $\lambda$.

This system of equations requires seven boundary conditions in order to be solved. At the inner boundary, we require
\begin{equation}
    \Psi = 0 \, 
\end{equation}
\begin{equation}
    \frac{\delta P}{P} = 0 \, ,
\end{equation}
and
\begin{equation}
   \delta \Phi = 0 \, ,
\end{equation}
%\begin{equation}
%    \frac{\delta T}{T} = 0 \, ,
%\end{equation}
%\begin{equation}
%    \frac{\delta L}{L_s} = 0 \, .
%\end{equation}
%\begin{equation}
%    \delta \Phi' = \frac{\ell+1}{r} \delta \Phi \, .
%\end{equation}
These ensure that $\delta T/T$, $\delta \rho/\rho$, and $\delta L/L_s$ are also zero at the center of the star. 

At the surface, we require the blackbody radiation condition
\begin{equation}
    \frac{\Delta L}{L_s} - 4 \frac{\Delta T}{T} - 2 \frac{\xi_r}{r} = 0 \, ,
\end{equation}
where $\Delta$ indicates a Lagrangian perturbation. Using $dL/dr=0$ at the surface, and the surface pressure boundary condition $\Delta P = 0$, which becomes $(P/\rho g r) \delta P/P = \xi_r/r \ll \delta P/P$, we can drop the last term, and this can be written as
\begin{equation}
    \label{eq:lsurf}
    \frac{\delta L}{L_s} - 4 \frac{\delta T}{T} + 4 \nabla \frac{\delta P}{P}= 0 \, ,
\end{equation}
where $\nabla = d \ln T/d \ln P$.
At the outer boundary we require a decaying potential perturbation:
\begin{equation}
    \delta \Phi' = -\frac{\ell+2}{r} \Phi \, .
\end{equation}
Additionally, the amplitude of the response can be chosen with a normalization condition at the surface, e.g., 
\begin{equation}
    \label{eq:bsurf}
    \Psi = \sqrt{G M^2} \, .
\end{equation}

Finally, we require another boundary condition at the surface that determines the amplitude of another variable (e.g., $\delta T/T$ or $\delta L/L_s$), relative to $\Psi$. This boundary condition will determine the amplitude of the surface flux perturbation and can be considered to be a measure of the strength of the magnetic forces within the star. It is similar to specifying the strength of the magnetic forces in barotropic stars with a $\beta$ parameter as described in \cite{duez:10} and in Section \ref{sec:con}. In practice, we set the luminosity perturbation at the outer boundary in order to determine an effective value of $\beta$. Setting $\delta L=0$ at the outer boundary would be equivalent to setting $\beta=0$.

We pause to note a few important points. First, the displacement vector $\bxi$ does not appear anywhere in our system of equations. Lagrangian perturbations cannot be calculated from this system of equations, except at the surface where $\xi_r/r = (P/\rho g r) \delta P/P$. Physically this arises from the fact that in a thermally relaxed system of uniform composition, there are an infinite number of combinations of radial and horizontal displacements $\xi_r$ and $\xi_\perp$ that could give rise to a given density perturbation $\delta \rho/\rho$. This is discussed further in Section \ref{sec:discussion}.

A second important point is that we have not imposed a surface boundary condition in which the poloidal or toroidal component of the magnetic field goes to zero. This is quite different from the fields studied in \cite{broderick:08} and \cite{duez:10}. A non-zero toroidal field requires a current to flow at the surface of the star, which is often assumed to be zero due to the vanishing density and temperature. Real stars, however, do not have zero temperature or density in their photospheres or coronae, which can support currents and toroidal fields \citep{Kochukhovetal2011,Shulyak:2007,Shulyak:2010}. Therefore, we do not impose $\Psi=0$ at the surface. A consequence of relaxing this condition is that $\lambda$ is no longer an eigenvalue, and the system of equations can now be solved for any value of $\lambda$.

Alternatively, one could argue that the radial component of the electric current (Equation \ref{eq:j}) should vanish at the surface of the star. This would require $B_r=0$ and hence $\Psi=0$ as discussed in Section \ref{sec:j}, and imposed by \cite{broderick:08} and \cite{duez:10}. This eighth boundary condition would transform our system of seven differential equations into an eigenvalue problem that could only be solved for certain combinations of $\lambda$ and the surface flux perturbation. We discuss this further in Section \ref{sec:solutions} and \ref{sec:discussion}.

\begin{figure*}
\includegraphics[scale=0.7]{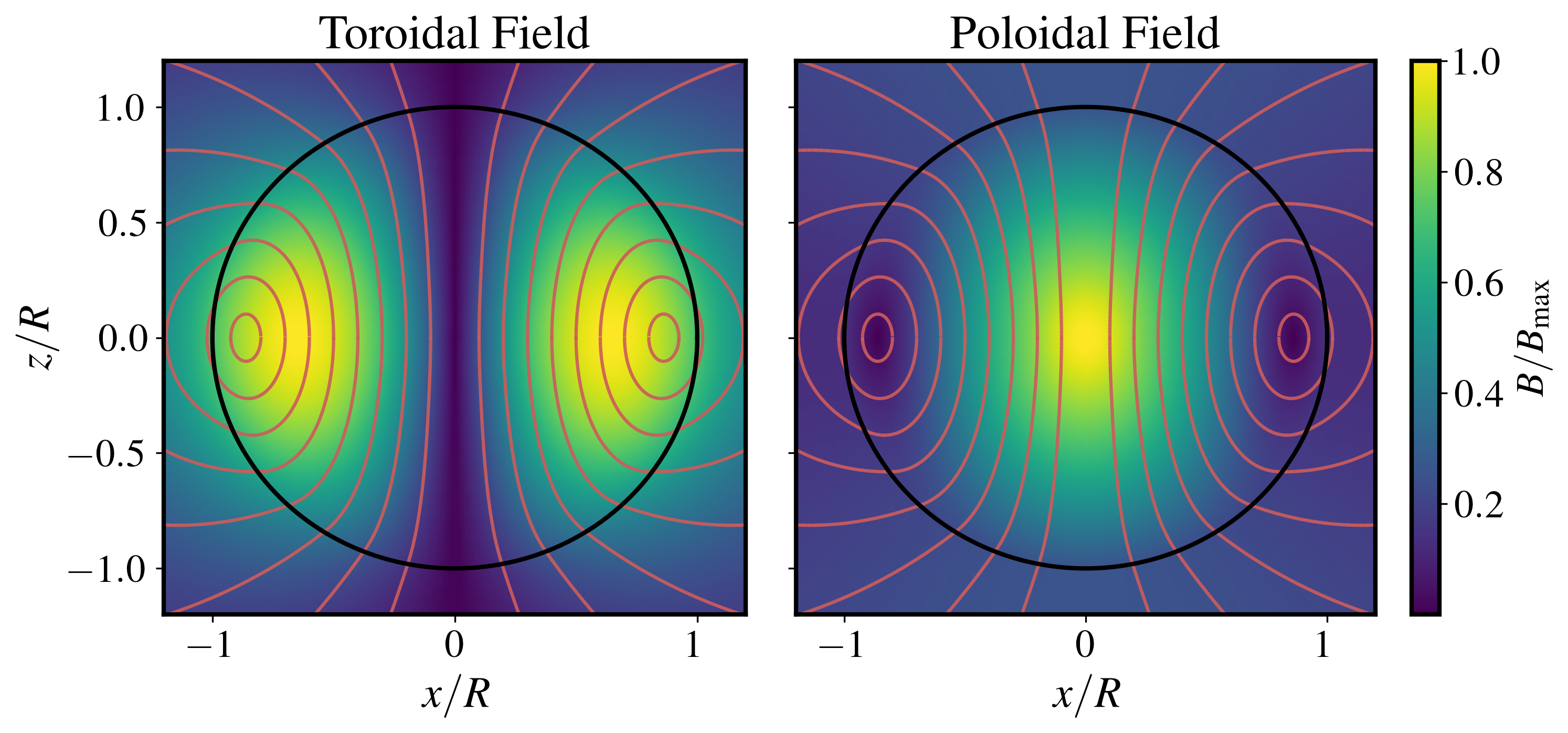}
\caption{\label{fig:magstream}
Meridional slices of a star, with color shading indicating the strength of the toroidal magnetic field (left) and poloidal magnetic field (right) normalized to the maximum magnetic field. Orange lines show magnetic field lines of the poloidal field. This model has magnetic helicity $\lambda^2=10$ and magnetic force $\beta=10$.}
\end{figure*}

\begin{figure}
\includegraphics[scale=0.37]{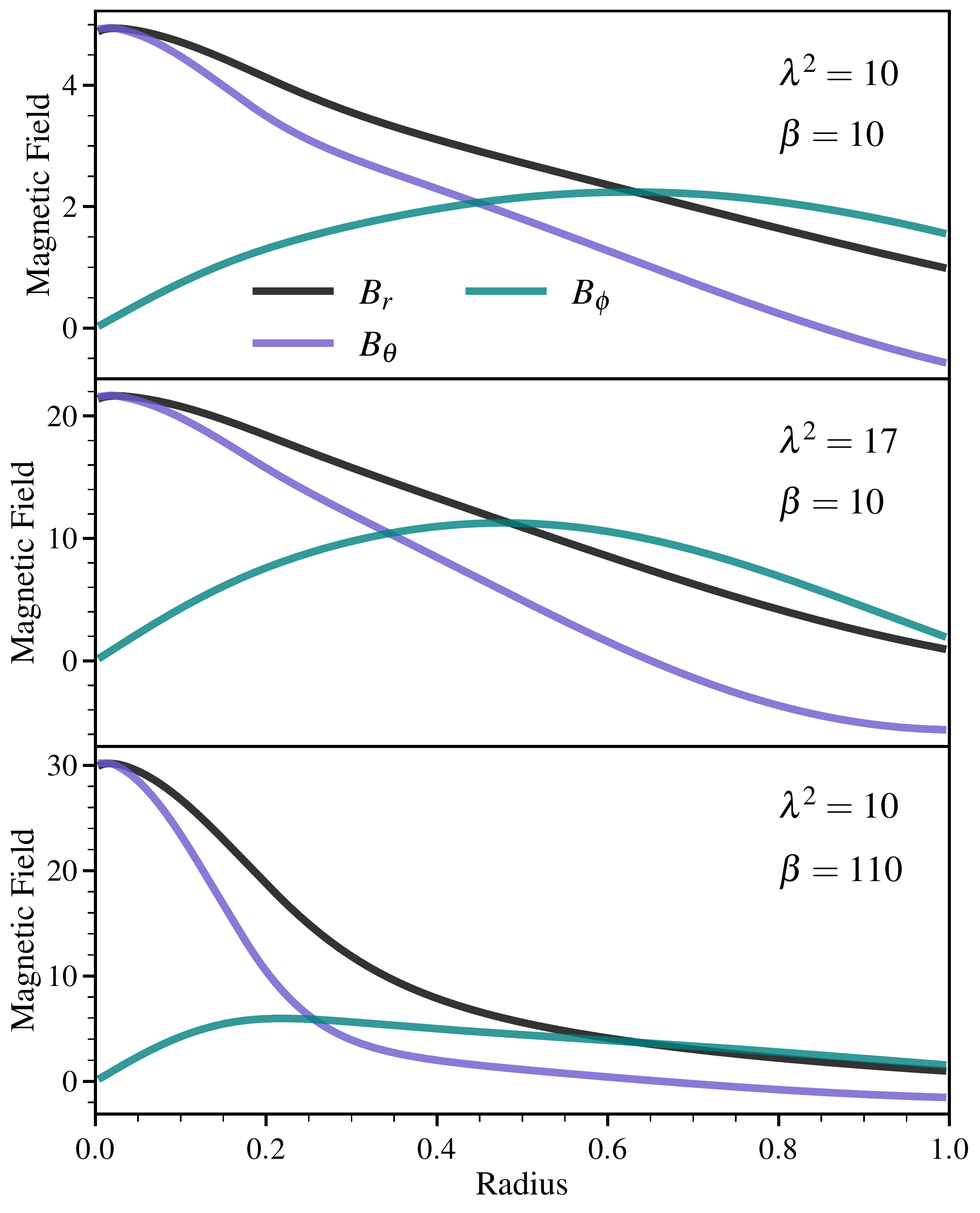}
\caption{\label{fig:magstrucB} Radial profiles of the $r$, $\theta$, and $\phi$-components of the magnetic field, scaled to the radial field at the surface, for models with $\beta=10$.
%{\bf as a function of $r$ normalized by the stellar radius $R$}.
Panels are labeled by their values of $\lambda^2$ and $\beta$. Note that the field is much more centrally concentrated for higher values of $\lambda$ or $\beta$.}
\end{figure}

\subsection{Convective Zones}
\label{sec:con}

In convective zones, the radiative diffusion equation no longer applies, and equations \ref{eq:dtdr} and \ref{eq:dldr} are not valid. Instead, we assume that the perturbation to the entropy is nearly zero, as convective heat flux will quickly smooth out any entropy gradients. We therefore have 
\begin{equation}
    \frac{d \ln T}{d \ln P} = \nabla_{\rm ad} = \frac{\Gamma_3 - 1}{\Gamma_1} \, .
\end{equation}
Perturbing this yields
\begin{equation}
\label{eq:dtb}
    \frac{\delta T}{T} - \frac{\Gamma_3 - 1}{\Gamma_1} \frac{\delta P}{P} = 0 \, .
\end{equation}
and
\begin{equation}
    \frac{\delta P}{P} - \Gamma_1 \frac{\delta \rho}{\rho} = 0 \, .
\end{equation}

Since the perturbations are barotropic, the system of equations \ref{eq:dpsidr}-\ref{eq:dpdr} simplifies, as discussed above. In this case, the system of equations reduces to that of \cite{duez:10}:
\beq
\label{eq:dpsidr2}
\frac{\partial}{\partial r} \Psi - \Psi' = 0 \, ,
\eeq
\beq
\label{eq:dpsipdr2}
\frac{\partial}{\partial r} \Psi' + \bigg(\frac{\lambda^2 }{R^2} - \frac{\ell(\ell+1)}{r^2} \bigg) \Psi + \beta r^2 \rho = 0 \, .
\eeq
Here, $\beta$ is a constant that determines the magnitude of the magnetic force. Comparison with equation \ref{eq:beta} shows that the pressure perturbation is
\begin{equation}
\label{eq:dpbeta}
    \frac{\delta P}{P} + \frac{\rho \delta \Phi}{P} = \frac{\beta \rho \Psi}{P} \, .
\end{equation}
%In our models, we solve equations \ref{eq:dpsidr2} and \ref{eq:dpsipdr2} subject to the boundary condition $\Psi=0$ at $r=0$, and with a normalization condition $\Psi=1$ at the outer edge of the convective zone. The inner boundary condition also enforces $\delta P/P=0$ and $\delta T/T=0$ as is necessary. The values of $\lambda$ and $\beta$ are constants that can be arbitrarily chosen.
Equations \ref{eq:dpsidr2} and \ref{eq:dpsipdr2} combine into a linear wave equation: 
\beq
\label{eq:ddpsidr}
\frac{\partial^2}{\partial r^2} \Psi + \bigg(\frac{\lambda^2}{R^2} - \frac{\ell(\ell+1)}{r^2} \bigg) \Psi + \beta r^2 \rho = 0 \, .
\eeq
%The value of $\lambda$ is a parameter that determines the magnetic helicity and also the radial wavenumber of the magnetic field. The value of $\beta$ determines the size of a non-homogeneous forcing term in the equation, i.e., the strength of the magnetic force. This is directly proportional to the perturbed pressure, density, and temperature produced by the magnetic field.

%{\bf In our numerical code, we do not enforce equation \ref{eq:dpsipdr2}, but rather enforce equation \ref{eq:dtb} within convective regions. Note that the luminosity perturbation $\delta L$ is not defined within convective regions. We set $\partial \delta L/\partial r=0$ in our code, but the actual luminosity perturbation is undetermined.}

\subsection{Linking Convective and Radiative Zones}
\label{sec:link}

In this work we only consider stars with convective cores and radiative envelopes. To solve for the magnetic field in the full star, we choose a value of $\lambda$ and a surface flux perturbation $\delta L/L$, which effectively sets the magnetic forces and the value of $\beta$ within the convective zone. Within the convective zone, we replace equation \ref{eq:dtdr} with equation \ref{eq:dtb}, and we replace equation \ref{eq:dldr} with $\partial \delta L/\partial r = 0$. However, the value of $\delta L$ is not defined in the convective zone, only within the radiative zone above it. In our solutions, we verify that equation \ref{eq:dpbeta} is approximately satisfied within the convective region. We then label our solutions by the corresponding value of $\beta.$

%$\beta$ and solve for the perturbed structure in the convective core as described above. Since the perturbations should be continuous across the radiative-convective boundary, the values of $\Psi$, $\Psi'$, $\delta P/P$, and $\delta T/T$ at the outer edge of the convective core are then inner boundary conditions for our solution in the radiative region, as described in section \ref{sec:rad}. In this case, the surface normalization boundary condition (equation \ref{eq:bsurf}) is not needed, because it is replaced by the normalization condition used in the convective core.

\subsection{Solving the Equations}

We solve the system of equations above in a stellar model generated with the MESA stellar evolution code \citep{paxton:11}. We choose a $M=3 \, M_\odot$ star at the start of the main sequence, with a radius of $R=2.1 \, R_\odot$, a surface temperature of $T_{\rm eff} = 11,800\,{\rm K}$, and a convective core boundary at $r/R = 0.13$. This model resembles typical magnetic Ap/Bp stars that are observed to harbor strong magnetic fields. Our model has nearly uniform stellar composition so that the equation of state (equation \ref{eq:eos}) is a good approximation.

We use a relaxation technique from Numerical Recipes \citep{press:07} to solve the system of equations in Section \ref{sec:equations}. We solve the equations on the same grid as the underlying MESA model. A good initial guess is often required in order reliably to converge to a solution. Spurious solutions (involving sudden jumps in the derivatives of $\delta P$ and $\Psi$) are often found, so caution is required. There may be other physical solutions that exist that we do not examine in this work.

\begin{figure}
\includegraphics[scale=0.36]{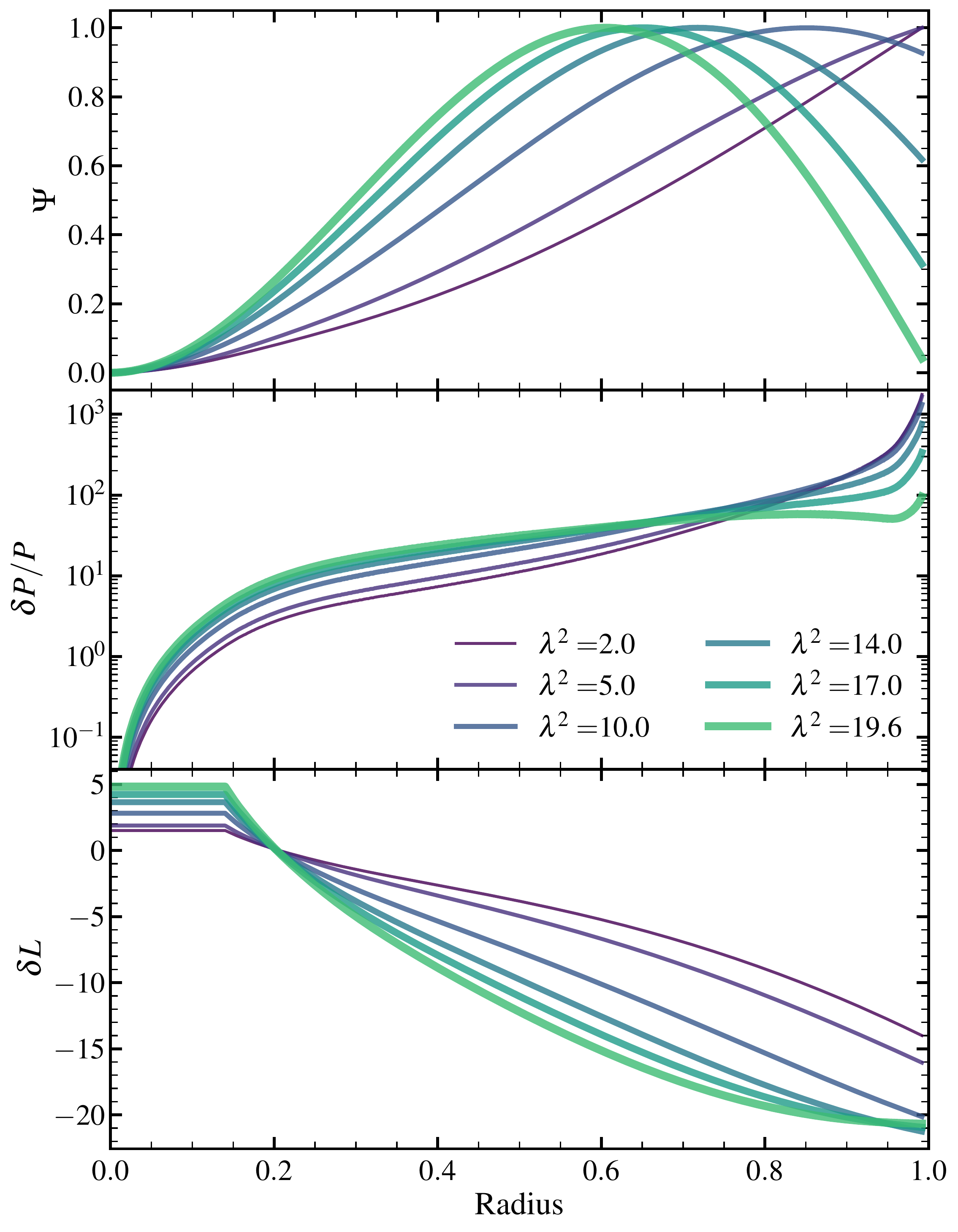}
\caption{\label{fig:magstrucb10} Profiles of the magnetic potential $\Psi$, relative pressure perturbation $\delta P/P$, and luminosity perturbation $\delta L$ (in units of $L_{\rm surf}$) as a function of radius for a model with $\beta=10$ and varying values of the helicity $\lambda$. All quantities are normalized so that $\Psi=1$ at its maximum. Stronger toroidal fields (larger $\lambda$) create more oscillatory magnetic fields, but the associated surface luminosity perturbation varies only slightly. The large values of $\delta P/P$ near the surface are discussed in \autoref{sec:nonlin}.}
\end{figure}

\begin{figure}
\includegraphics[scale=0.36]{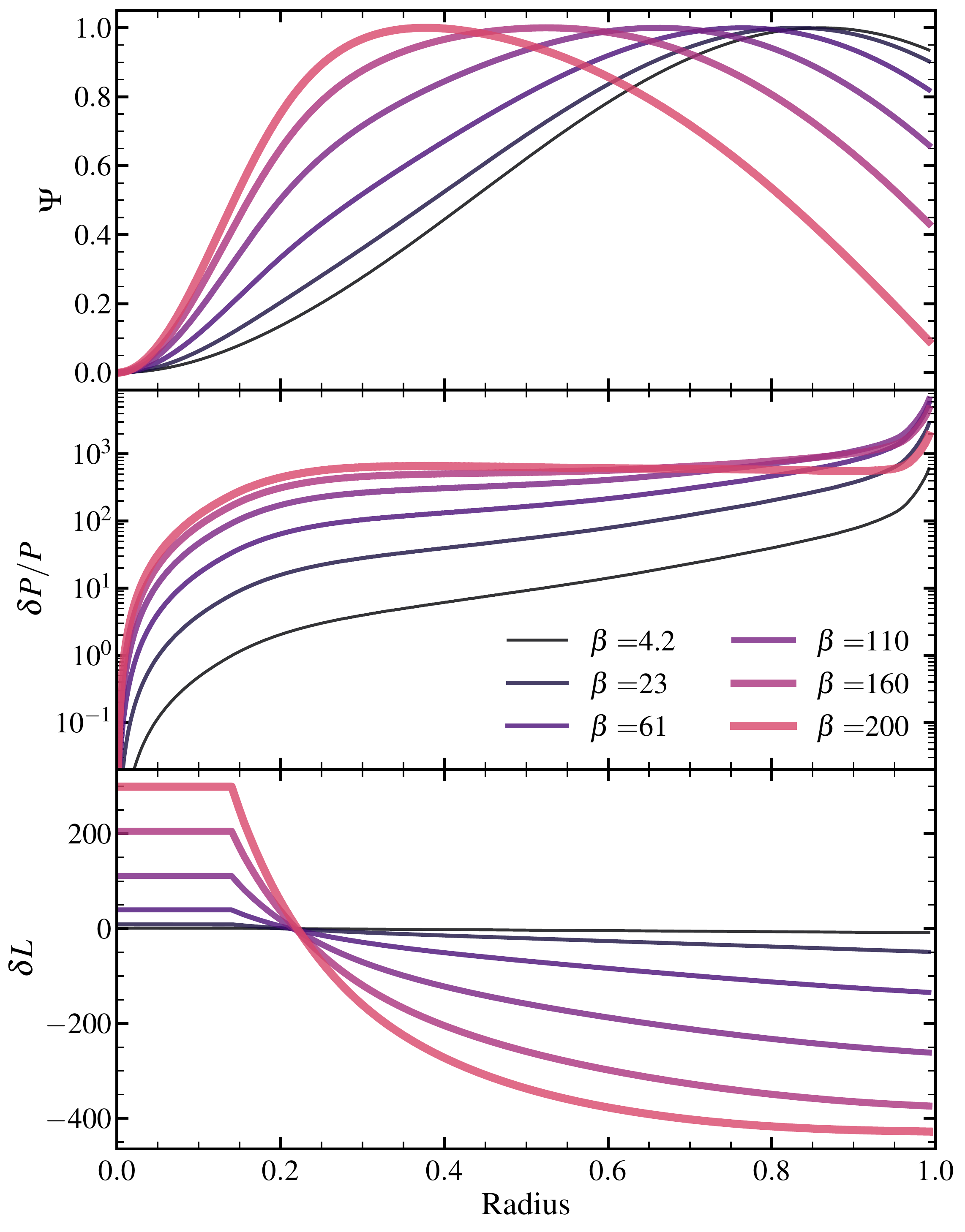}
\caption{\label{fig:magstrucl10} Same as \autoref{fig:magstrucb10}, but for a model with $\lambda^2=10$ and varying values of $\beta$. Larger magnetic forces (higher values of $\beta$) create larger surface luminosity perturbations and more oscillatory magnetic fields.}
\end{figure}

\begin{figure}
\includegraphics[scale=0.36]{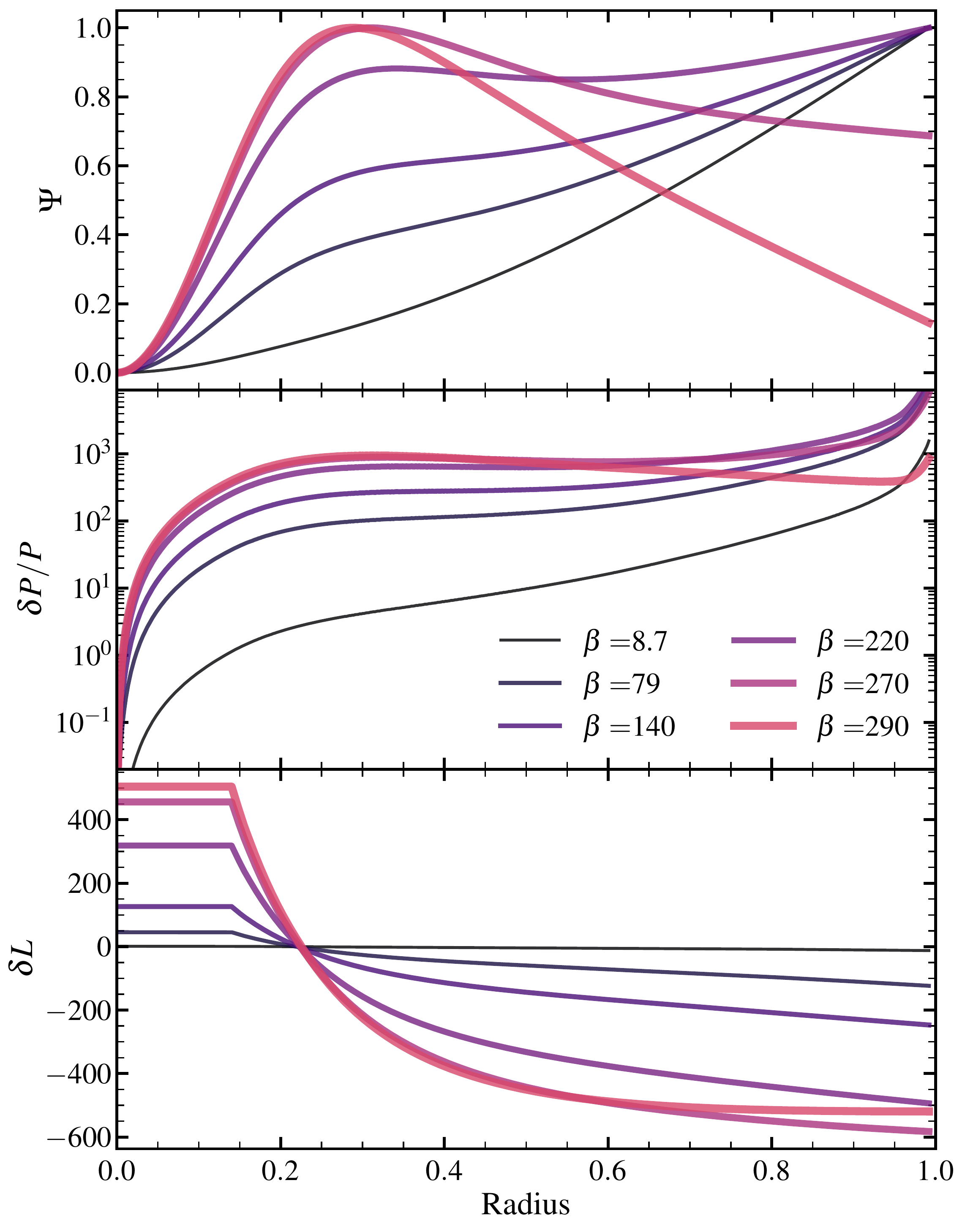}
\caption{\label{fig:magstrucl2} Same as \autoref{fig:magstrucl10}, but for a model with $\lambda^2=2$. Although the magnetic structure is somewhat different, the surface luminosity perturbations are similar to the case in \autoref{fig:magstrucl10} with stronger toroidal fields.}
\end{figure}

\section{Solutions}
\label{sec:solutions}

In our models, both $\lambda$ and $\beta$ are free parameters. Physically, $\lambda$ represents the magnetic helicity which is determined by the dynamo process that created the field, and which is roughly conserved during subsequent turbulent relaxation \citep{Braithwaite2008,duez:10}. The lowest energy stable field configurations have $\lambda \sim 1$, so $\lambda$ values in the range of 1-10 are reasonable expectations for real stars, though larger values are possible. The value of $\beta$ determines the relative strength of magnetic forces as described in the previous section.
%In models without a convective core, $\beta$ does not appear in the equations, but the relative values of gas perturbations (e.g., $\delta P$, $\delta T$, etc.) are still determined by the strength of the magnetic forces, and can be set by adjusting the values of these quantities at the outer boundary.

\autoref{fig:magstream} and \ref{fig:magstrucB} show the magnetic field configuration for a model with $\lambda^2=\beta=10$. The poloidal field is largest at the center of the star, where its field strength is roughly four times larger than the surface value. The toroidal field is largest at $r/R \approx 0.7$. We shall see below that $\beta=10$ is small such that the magnetic field is similar to a force-free field. For this value of $\lambda$, there is a null point (where $B_\theta=0$ and the field lines converge to closed loops at the equator) at $r/R \simeq 0.85$.
%Higher values of $\lambda$ or $\beta$ have null points deeper in the star.
Both the poloidal and toroidal fields extend above the surface of the star, where the field is force-free but is not current-free. 

\autoref{fig:magstrucb10}, \ref{fig:magstrucl10}, and \ref{fig:magstrucl2} show the magnetic potential $\Psi$, pressure perturbation $\delta P/P$, and luminosity perturbation $\delta L/L$ as a function of radius. In \autoref{fig:magstrucb10}, each curve has a different value of $\lambda$. Higher values of $\lambda$ cause more oscillatory variation of $\Psi$, as can be seen in equation \ref{eq:ddpsidr} where the radial wavenumber of $\Psi$ is roughly $\sqrt{\lambda^2/R^2 - \ell(\ell+1)/r^2}$. Higher values of $\lambda$ push the null point and toroidal field maximum deeper into the star, and also cause the central field strengths to become larger relative to the surface field strength, as shown in \autoref{fig:magstrucB}. In this stellar model, a value of $\lambda \! \simeq \! 20$ is the first value of $\lambda$ for which $\Psi=0$ at the surface. From equation \ref{eq:dpsipdr} we see that $\Psi=0$ requires $\delta P=0$, so the Eulerian pressure perturbation is always zero where the radial component of the field is zero. 

For larger values of $\lambda$ or $\beta$, the values of $\Psi$ and $\delta P$ approach zero somewhere within the model. When this happens, the numerical solutions exhibit strange behavior. The values of $\Psi'$ and $\partial \delta P/\partial r$ sometimes exhibit discontinuous jumps at the zero-crossings of $\Psi$. However, the radial magnetic field, pressure perturbation, and temperature perturbations are all continuous across these zero-crossings (only their derivatives are discontinuous). It is unclear if these solutions are physical or numerical artifacts. Physically, these solutions would exhibit a discontinuity in the $\theta$-component of the magnetic field, an associated current sheet, and a discontinuity in both the magnetic force and the pressure force. Because the physicality of these solutions is unclear and they also sometimes cause numerical convergence problems, we will not investigate them further in this work. More highly oscillatory solutions also represent higher energy states \citep{broderick:08} and may be less likely to exist in stars that have relaxed to a minimum energy state. 

In \autoref{fig:magstrucb10}, the relatively small value of $\beta=10$ means that the pressure and density terms in equation \ref{eq:dpsipdr} are negligible relative to the magnetic terms, and the field is nearly force-free. The magnetic solutions are thus similar to the force-free solutions of \cite{broderick:08}. In this limit, the perturbed pressure, temperature, etc. have a value that is proportional to $\beta$, with a radial dependence that is determined only by $\lambda$ and the structure of the star. This can be seen from equation \ref{eq:dpbeta}, such that the value of $\delta P$, $\delta T$, and $\delta L$ are roughly proportional to $\beta$ at the radiative convective interface, and hence within the bulk of the radiative zone. It is demonstrated in \autoref{fig:magstrucl10} and \ref{fig:magstrucl2}, where the pressure and luminosity fluctuations have similar profiles and increase linearly with $\beta$, as long as $\beta \lesssim 100$. Hence, a wide range in surface temperature and flux variations are possible for a given surface field, depending on the strength of the magnetic forces, parameterized by $\beta$ in these models. 

For larger values of $\beta$, however, the pressure/density terms in equation \ref{eq:dpsipdr} become comparable to the magnetic terms. Physically this means that gas pressure forces begin competing with magnetic forces, altering the profile of the magnetic field. Large values of $\beta$ have a similar effect to larger values of $\lambda$, causing more oscillatory behavior of the magnetic potential $\Psi$.
%Figures \ref{fig:magstrucl10} and \ref{fig:magstrucl2} show the behavior of $\Psi$, $\delta P/P$, and $\delta L$ for increasingly large values of $\beta$.
The surface pressure and luminosity perturbations reach a maximum when $\beta \sim 200$, and decrease at larger values. This is because $\Psi$ (and hence $\delta P$) become smaller near the surface at large values of $\beta$ as gas pressure forces start backreacting on the magnetic field profile. The magnitude of $\beta$ needed to have a large influence on the field profile can be seen from equation \ref{eq:ddpsidr}: $\beta$ must be large enough for the last term to be comparable to the second term. Deep within the star, this requires $\beta \sim \ell(\ell+1) \Psi/(\rho r^4)$, which typically has a value of a couple hundred for our stellar model and normalization.

\begin{figure}
\includegraphics[scale=0.38]{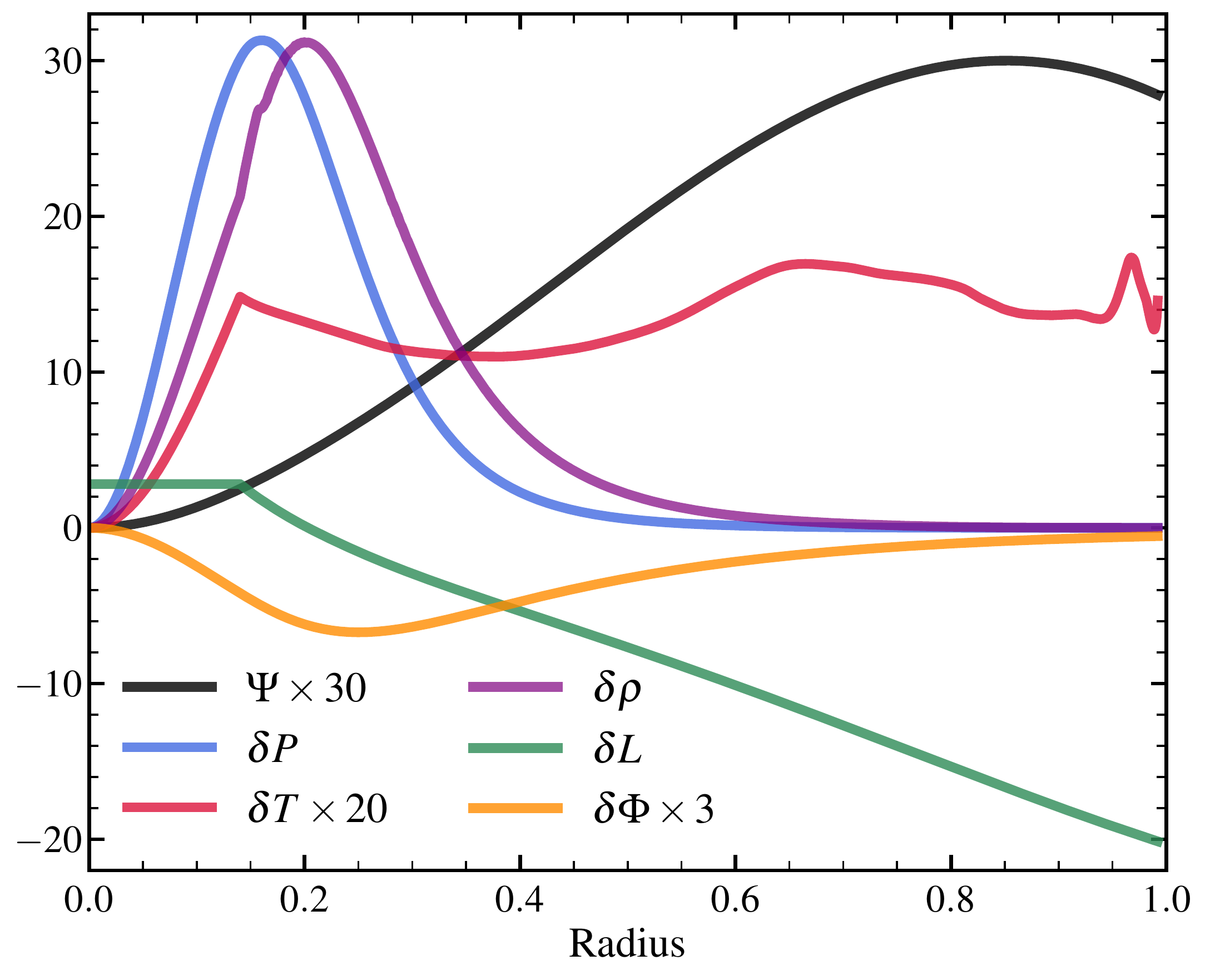}
\caption{\label{fig:magstrucl10b10} The magnetic potential $\Psi$ (units of $\sqrt{GM^2}$), pressure perturbation $\delta P$ (units of $GM^2/R^4$), temperature perturbation $\delta T$ (units of $T_{\rm cen}$), density perturbation $\delta \rho$ (units of $M/R^3$), luminosity perturbation $\delta L$ (units of $L_{\rm surf}$), and gravitational potential perturbation $\delta \Phi$ (units of $GM/R$) for a model with $\lambda^2=10$, $\beta=10$. The kinks at $r=0.13$ occur at the boundary of the convective core.}
\end{figure}

From \autoref{fig:magstrucb10}-\ref{fig:magstrucl2}, we see that $\delta P/P$ typically approaches very large values near the surface of the star, because $P$ reaches very small values. However, the value of $\delta P$ approaches very small values $\delta P \simeq \rho g \xi_r$ near the surface, such that the Lagrangian pressure perturbation $\Delta P = \delta P - \rho g \xi_r$ smoothly approaches zero at the surface. The value of $\delta P/P$ thus approaches $\delta P/P \simeq \xi_r/H$ at the star's surface, which becomes large as the pressure scale height $H$ becomes small. Although $\delta P/P$ peaks near the surface, the value of $\delta P$ peaks at radii of $r/R\simeq 0.2$, as can be seen in \autoref{fig:magstrucl10b10}. The values of $\delta \rho$ and $\delta \Phi$ show similar behavior.

In contrast, the value of $\delta T$ increases within the convective zone and then maintains a roughly constant profile throughout the star. The derivative of $\delta T$ has a discontinuity at the convective interface due to the change in structure and energy transport mechanism. Within the convective core, $\delta T$ is directly proportional to $\delta P$ which is proportional to $\Psi$ (equations \ref{eq:dtb} and \ref{eq:dpbeta}), but in the radiative zone $\delta T$ is determined by the thermal equilibrium conditions (equations \ref{eq:dtdr} and  \ref{eq:dldr}). This causes oscillatory variations in $\delta T$ in the outer layers of the star, due to variations in the thermal diffusivity $\chi$ caused by opacity variations in partial ionization zones near the surface. 

Nevertheless, the luminosity perturbation $\delta L$ always changes smoothly and gradually throughout the star. Physically this occurs because sudden changes in $\delta L$ would be smoothed out by radiative diffusion. Mathematically this can be seen from equation \ref{eq:dldr}, because the values of $\chi T r/L$ and $d \ln L/d \ln r$ become very small near the surface of the star, preventing sudden variation in $\delta L$, despite large values of $\delta P/P$ and $\delta T/T$. Interestingly, in all of our models, the luminosity perturbation switches sign at $r/R \simeq 0.23$ within the radiative region above the convective core. 

%The scale of $\delta P$, $\delta T$, and $\delta L$ are all roughly proportional to $\beta$ for small values of $\beta$ within the convective zone. 

\subsection{Surface Flux Perturbation}

All of our models have negative surface luminosity perturbations $\delta L_2$ as shown in the figures. However, the quadrupolar component of the physical response (see equation \ref{eq:decomp}) is $\delta L = \delta L_2 (1/3-\cos^2 \theta)$. Hence, a negative value of $\delta L_2$ translates to a positive flux perturbation at the magnetic pole and a negative flux perturbation at the magnetic equator, as shown in \autoref{fig:magstruc}. This is consistent with the heuristic idea that strong magnetic pressure at the star's pole causes the gas pressure and density to be smaller, allowing us to see deeper into the star such that magnetic spots are brighter \citep{Cantiello:11}. However, it is also possible for our models to produce a negative flux perturbation at the magnetic pole if we consider negative values of $\beta$, so in principle it is possible for the magnetic pole to be either hot or cool. Below we argue that a hot magnetic pole is more likely.

The value of the surface flux perturbation in our models is $\delta L/L \sim \beta \Psi_{\rm max}^2$, where $\Psi_{\rm max}$ is the maximum value of $\Psi$ reached within the model. Since we have normalized $\Psi_{\rm max}$ to units of $\sqrt{GM^2}$, this implies
\begin{equation}
    \frac{\delta L}{L} \sim \frac{\beta}{4 \pi} \frac{B_{\rm max}^2 r_{\rm max}^4}{GM^2} \, ,
\end{equation}
where $B_{\rm max}$ and $r_{\rm max}$ are the magnetic field and radius where $\Psi$ peaks. In our models, this roughly translates to  
\begin{equation}
\label{eq:dlsurf}
    \frac{\delta L}{L} \sim 10^{-11} \beta \bigg(\frac{B_{\rm surf}}{1 {\rm kG}}\bigg)^2 \bigg(\frac{R}{R_\odot}\bigg)^4 \bigg(\frac{M}{M_\odot}\bigg)^{-2} \, ,
\end{equation}
where $B_{\rm surf}$ is the magnetic field at the star's surface. Clearly this is too small to be detected in main sequence stars unless extremely large values of $\beta$ are assumed.
Equation \ref{eq:dlsurf} is consistent with a naive estimate from the von Zeipel theorem applied to the star's interior.
%One might expect the flux perturbation due to centrifugal distortion is $\delta F/F \sim f_{\rm cen}/g$, where the centrifugal force is $f_{\rm cen} = \Omega^2 r$, with $\Omega$ the rotation rate.
The smoothly varying magnetic fields of our models produce magnetic forces of order $f_{\rm mag} \sim \beta B^2/(\rho r)$. Using the average stellar density $\rho \sim M/R^3$ and gravitational force $g \sim GM/R^2$ yields the von Zeipel estimate of $\delta F/F \sim f_{\rm mag}/g \sim \beta B^2 R^4/GM^2$.
%We have used the fact that the interior field is usually several times larger than the surface field, canceling factors of $\sim 4 \pi$. 

\begin{figure}
\includegraphics[scale=0.37]{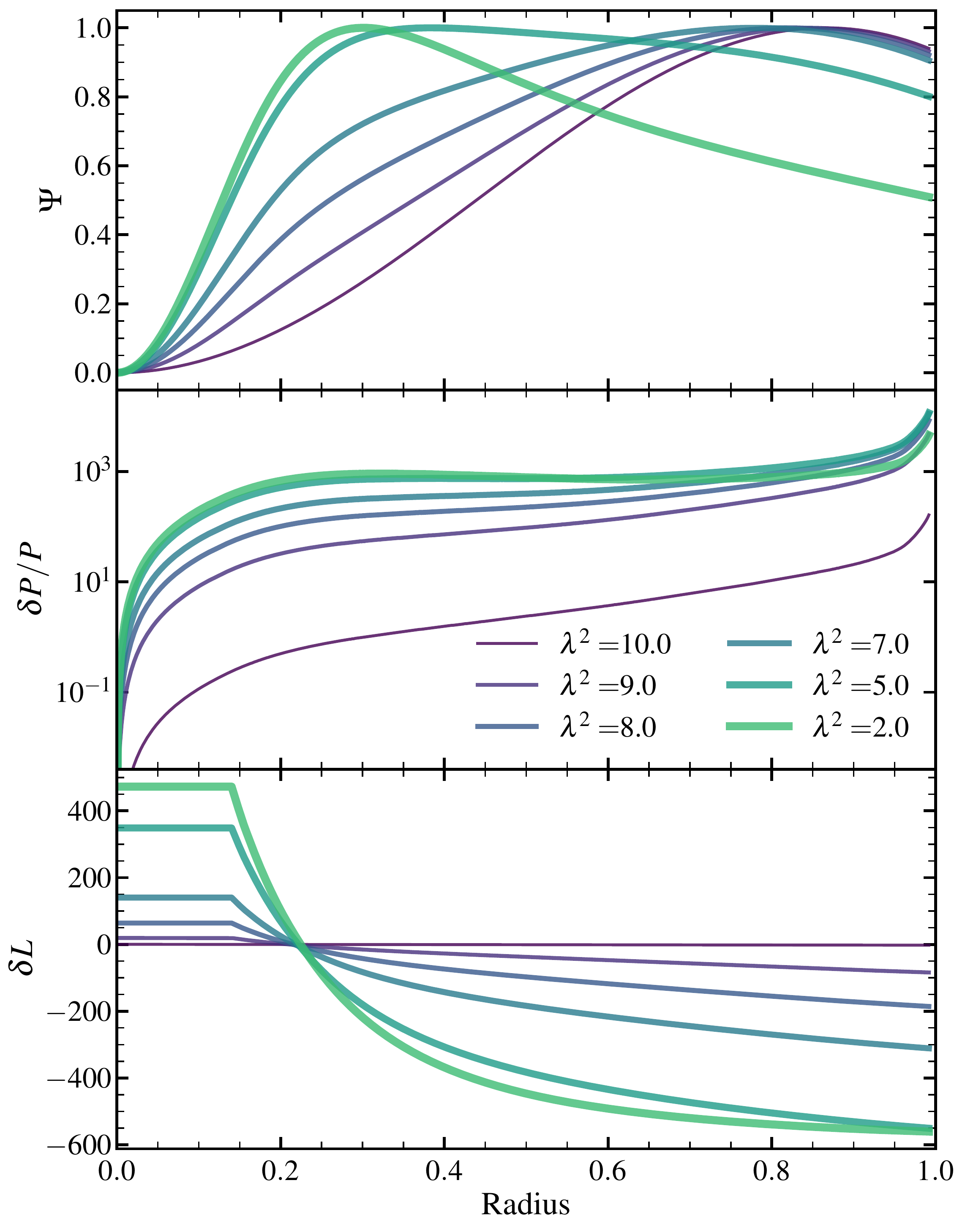}
\caption{\label{fig:magstrucpotential} Magnetic structures for models whose poloidal component has a potential profile ($\Psi \propto r^{-\ell}$) at the outer boundary, for various magnetic helicities $\lambda$. This outer boundary condition effectively sets the strength of magnetic forces, $\beta$. Configurations similar to these may be more likely to exist in real stars.}
\end{figure}

\begin{figure}
\includegraphics[scale=0.37]{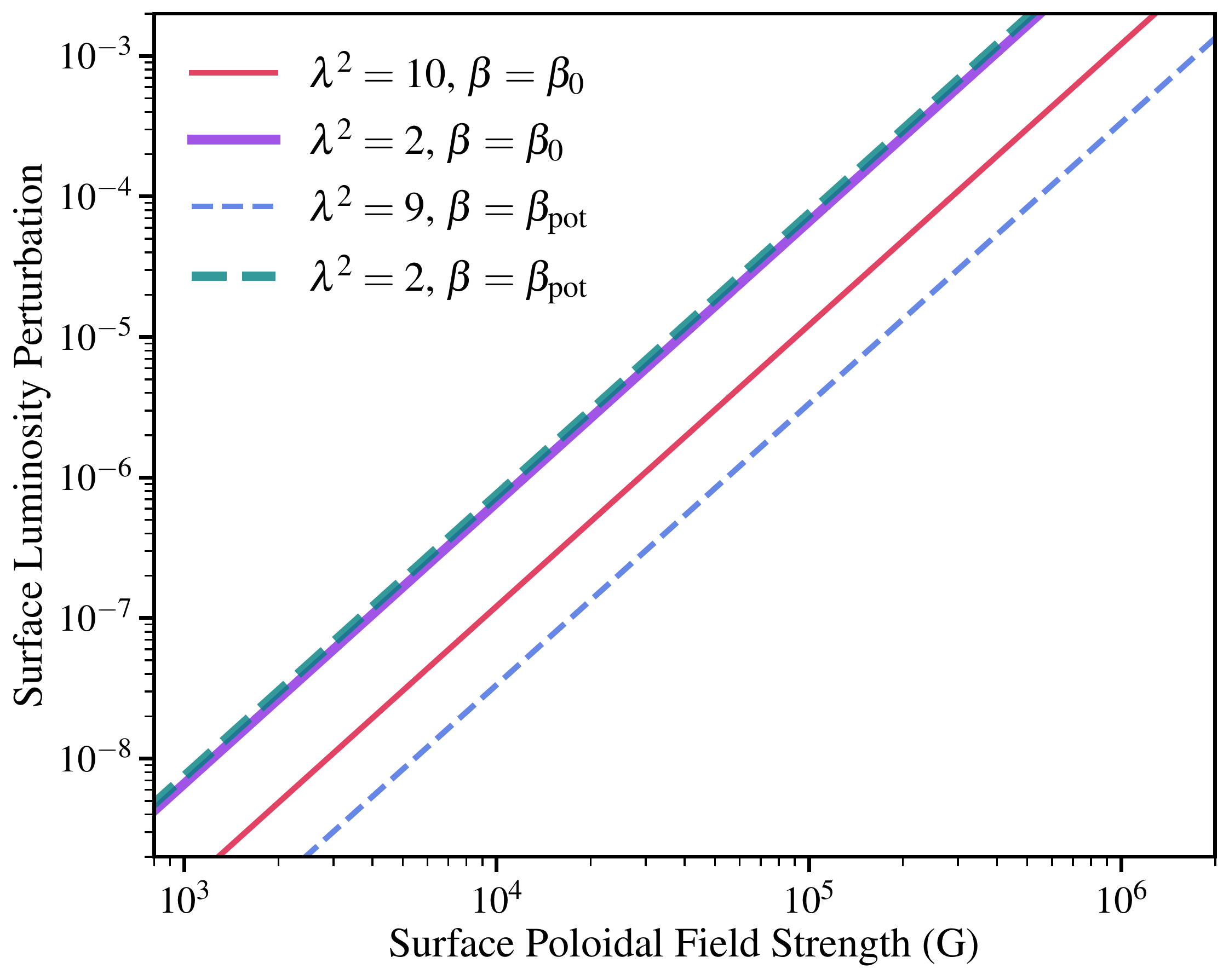}
\caption{\label{fig:magstrucvis} The surface luminosity perturbation as a function of surface magnetic field strength for models with $\beta=\beta_0$, such that the radial component of the magnetic field and electric current vanish at the surface. We also plot luminosity perturbations for $\beta=\beta_{\rm pot}$ such that the poloidal field has a potential profile at the surface. In either case, very strong magnetic fields are required for detectable luminosity perturbations in main sequence stars. Observed photometric variations of magnetic stars likely arise from variations in the emergent spectrum rather than variations in the bolometric flux.}
\end{figure}

\subsection{Most likely field configurations}

As mentioned in Section \ref{sec:j}, the radial current $j_r$ may be expected to be close to zero near the surface of the star so that current does not flow into the star's low-density corona. From equation \ref{eq:j}, the only way this can happen is when $\lambda^2=0$ or $\Psi=0$ at the surface of the star. If there are near-surface toroidal fields, $\Psi$ (and hence $B_r$) must be zero to have a non-zero current flowing out of the star. Hence there is a critical value $\beta_0$ such that $\Psi=0$ at the star's surface. \autoref{fig:magstrucl10} indicates this occurs at $\beta_0 \! \simeq \! 200$ and $\delta L/L \sim -400$ for $\lambda^2=10$ in our stellar model, or $\beta_0 \simeq 300$ and $\delta L/L \sim -500$ for $\lambda^2=2$ (\autoref{fig:magstrucl2}). Hence, those solutions are arguably the most physically probable amongst the possibilities shown here. Firef{fig:magstrucvis} shows the predicted surface flux variations for $\beta=\beta_0$. Even though the large values of $\beta$ increase the surface flux perturbation above the naive estimate discussed above, it is too small to be detected with current instruments. 

Another possibility is that the value of $\lambda^2$ is not constant within the star due to dissipative effects near the surface. In that case, one may expect the value of $\lambda$ to decrease near the star's surface, allowing the radial current to vanish and still have a large value of $\Psi$ and hence a non-zero radial magnetic field. If the electric currents decrease to zero near or above the surface, the field must approach a potential configuration. Potential force-free fields have $\partial \Psi/\partial r^2 = \ell(\ell+1)\Psi/r^2$, and an outwardly decreasing field has $\Psi \propto r^{-\ell}$ and hence $B \propto r^{-(\ell+2)}$, i.e., a dipole field for our assumed value of $\ell=1$. To model this case, we set the outer boundary condition to $\partial \Psi/\partial r = - \ell \Psi/r$. The fixed slope at the outer boundary effectively sets the value of $\beta$, which we refer to as $\beta_{\rm pot}$.

\autoref{fig:magstrucpotential} shows the results of this calculation for several values of $\lambda^2$. The required values of $\beta_{\rm pot}$ are typically similar to (but smaller than) $\beta_0$, i.e.,  values on the order of a hundred. For $\lambda^2\simeq 10$, a nearly force-free field satisfies the potential outer boundary condition, such that $\beta_{\rm pot}\simeq0$. For larger values of $\lambda$, the required $\beta_{\rm pot}$ is negative. However, we find those solutions almost always have a zero-crossing somewhere in the star, and cause numerical problems, so we do not investigate them further in this work. We note that they would likely entail surface flux perturbations of the opposite sign, i.e., a cool magnetic pole and hot magnetic equator. \autoref{fig:magstrucvis} shows the surface luminosity perturbations for $\beta=\beta_{\rm pot}$, which are a similar magnitude to those with $\beta=\beta_0$, and not detectable for observed magnetic field strengths.

\subsection{Shape of Distorted Star}

The sign of the pressure/density perturbations near the surface of the star determine wither it becomes oblate (smaller photospheric radius at magnetic pole) or prolate (larger radius at magnetic pole). As discussed above, we believe negative Eulerian pressure perturbations at the magnetic pole are most likely, which means the surface displacement $\xi_{r} = \delta P/(\rho g)$ is also negative at the magnetic pole. Hence, magnetically distorted stars are likely to be oblate. Heuristically, this is consistent with the idea that high magnetic pressure at the star's pole (or magnetic tension near the equator) squeezes the star into a flattened shape.

The star's ellipticity is 
\begin{align}
    \varepsilon &= \frac{\xi_{r,{\rm surf}}}{R} \nonumber \\
    &= \frac{\delta P}{P} \frac{H}{R} \nonumber \\
    &\sim \beta \times 10^{-10} \bigg(\frac{B_{\rm surf}}{1 {\rm kG}}\bigg)^2 \bigg(\frac{R}{R_\odot}\bigg)^4 \bigg(\frac{M}{M_\odot}\bigg)^{-2} \ \, .
\end{align}
Similar to the flux perturbations, this is too small to detect. The quadrupole moment of the star is 
\begin{align}
    Q &= \frac{\int \delta \rho r^4 Y_{20}^* d \Omega dr}{M R^2} \nonumber \\
    &\sim \beta \times 10^{-12} \bigg(\frac{B_{\rm surf}}{1 {\rm kG}}\bigg)^2 \bigg(\frac{R}{R_\odot}\bigg)^4 \bigg(\frac{M}{M_\odot}\bigg)^{-2} \ \, .
\end{align}
The quadrupole moment of the star is even tinier than other perturbations because the density perturbation is largest near $r\sim0.2$, providing a small lever arm for the quadrupole moment.

\subsection{Ohmic Heating and Poynting Flux}

Our models ignore non-ideal effects such as Ohmic heating that will inevitably cause the field to decay. As long as these effects are small, our approximations are suitable. We first investigate the effect of Ohmic heating in our models. The heating rate per unit volume is
\begin{equation}
\rho \epsilon_{\rm Ohm} = 4 \pi \eta j^2
\end{equation}
where the current density $j$ is given in equation \ref{eq:j}. Hence the heating rate is
\begin{equation}
\label{eq:heat}
    \rho \epsilon_{\rm Ohm} = \eta \frac{\lambda^2}{4 \pi R^2} \big( B_r^2 + B_\theta^2 \big) + \frac{\eta}{r^2 \sin^2 \theta} (\Delta^* \Psi)^2 \, .
\end{equation}
Using equations \ref{eq:br}, \ref{eq:bp}, and manipulating equation \ref{eq:hydrotheta2}, this can be expressed 
\begin{align}
\label{eq:heat2}
    \rho \epsilon_{\rm Ohm} &= \frac{\eta \sin^2 \theta }{r^2} \bigg[ \frac{-4}{r^2 R^2} \Psi^2 + \frac{1}{R^2} \Psi'^2 \nonumber \\
    &+ \bigg( \frac{r^2 (\delta P + \rho \delta \Phi)}{\Psi} + \frac{\lambda^2}{R^2} \Psi \bigg)^2 \bigg] \, .
\end{align}
The angular dependence is $\sin^2 \theta$, and we have removed a spherically symmetric term since we are only interested in the non-spherical component. 

Equation \ref{eq:heat2} can be compared to other sources of heat generation and/or heat diffusion in equation \ref{eq:dldr}. We find that the Ohmic heating term is roughly three orders of magnitude smaller than the heat diffusion term (second term in equation \ref{eq:dldr}) for all the models shown in this paper. Hence, Ohmic heating is irrelevant for these models.
%However, the magnetic diffusivity and Ohmic heating rate rises sharply towards the star's surface and could become large near and above the photosphere. 

Likely a more important effect is that the Ohmic diffusion time $t_{\rm Ohm} = r^2/\eta$ drops sharply near the surface of the star, falling to $\sim 1 \, {\rm Gyr}$ near the surface of our model. The diffusion time scale is even shorter in the atmosphere of the star and could be shorter than the star's lifetime. Consequently, the fields will dissipate or change their morphology within the star's atmosphere until they approach a current-free (and force-free configuration). This will bend the field lines and hence create magnetic forces within the star until it approaches a new (quasi)-equilibrium.

Another way of seeing this is by examining the Poynting flux (see \citealt{duez:10c}):
\begin{equation}
    F_{\rm poy} = \bnab \cdot \big( \eta {\boldmath F}_{\rm mag} \big)
\end{equation}
where ${\boldmath F}_{\rm mag} = (\bnab \times {\bf B}) \times {\bf B}/4 \pi$ is the Lorentz force. The Poynting flux represents the decrease of electromagnetic energy per unit volume, and it has the same order of magnitude as the Ohmic heating rate. As the fields dissipate, their energy density changes until the Poynting flux is nearly zero, which happens on Ohmic diffusion time scales.
\cite{landstreet:87} discusses the importance of magnetic forces arising due to Ohmic dissipation, finding they are likely negligible for main sequence stars. However, over long time scales, the morphology of the magnetic field will be altered away from the solutions we have computed, which could also affect the emerging luminosity perturbations.

\subsection{Linearity and near-surface effects}
\label{sec:nonlin}

Since the value of $\delta P/P$ can become very large near the surface of the star (\autoref{fig:magstrucl10}-\ref{fig:magstrucpotential}), non-linear effects may also start to be important in the near-surface layers. In our models, the $\delta P/P$ eigenfunction is 1-2 orders of magnitude larger than the surface luminosity perturbation $\delta L/L$, which is plotted as a function of surface field strength in \autoref{fig:magstrucvis}. Hence, $\delta P/P$ remains very small except for unphysically large surface field strengths $B \gtrsim 10^5 \, {\rm G}$.

However, we placed the outer boundary at a radius of $r/R \simeq 0.99$ in our calculations. This allowed us to avoid uncertainties associated with the response of near-surface convective layers above our outer boundary. The value of $\delta P/P$ likely continues to increase towards the surface (because $P$ drops sharply), so it is possible that non-linear effects start to become important near the surface. It is unlikely that this affects the luminosity perturbation, which changes smoothly with radius and cannot be greatly affected in the near-surface layers. Hence, the magnetic field profiles and luminosity perturbations that we calculate are probably robust, but the near-surface pressure and temperature profiles could be affected by these surface effects. In principle, this could affect the spectrum of the star, which is sensitive the atmospheric temperature profile.

\section{Discussion}
\label{sec:discussion}

\subsection{Stability of Equilibrium}

The magnetic field configurations we have computed are in hydrostatic and radiative equilibrium, but we have not investigated whether these equilibria are stable or unstable. \cite{braithwaite:09} (see also \citealt{akgun:13,becerra:22b}) showed that stable magnetic equilibria require $E_{\rm tor} \gtrsim 0.25 E_{\rm pol}$, where $E_{\rm tor}$ and $E_{\rm pol}$ are the toroidal and poloidal magnetic field energies. All of our models satisfy this criterion (they typically have $E_{\rm tor}/E_{\rm pol} \sim 1$) except for the $\lambda^2=2$ models. They also fall below the upper limit for stability, $E_{\rm mag} \! \lesssim \! (1/10) GM^2/R$ for surface field strengths less than $\sim \! 1 \, {\rm MG}$. Nonetheless, it is possible that some of our models are unstable, which should be examined in future work using 3D numerical simulations \citep{duez:10b,Kaufman:2022}.

\subsection{More Realistic Field Configurations}
A limitation of our work is the assumption of a field with purely dipole structure, whereas real fields likely have a more complicated angular structure that changes with radius. As discussed above, real fields likely have vanishing current above the photosphere which (according to our solutions, equation \ref{eq:j}) require vanishing $B_r$ or vanishing $\lambda$. The first conflicts with observations of real stars \citep{LandstreetMathys2000,oksala:18,Shultz:2019} while the second implies purely poloidal fields which are well known to be unstable \citep{markeytayler:73}. It is likely that a real star has a more complicated angular and radial field dependence, such that the electric currents vanish in the near-vacuum outside the star. In these configurations, the toroidal magnetic field vanishes on magnetic field lines that penetrate the surface of the star \citep{lyutikov:10}.

These sorts of configurations have been computed near the surface of a star in
\cite{raadu:71,milsom:76}, or in the interior for parameterized field configurations \citep{lyutikov:10,akgun:13,becerra:22b}. However, these configurations are not in thermal equilibrium and therefore not stable over thermal time scales. 
Computing such fields in the bulk of a star and accounting for thermal and hydrostatic equilibrium will require the solutions of partial differential equations, which is beyond the scope of this work. We suspect that such configurations (which appear qualitatively similar to those we compute) will alter our results by a factor of order unity, but will not greatly change any of our conclusions.

\subsection{Estimating the Perturbed Surface Flux}

Our solutions which map onto $B_r=0$ or $B_r \propto r^{-(\ell+2)}$ may resemble more realistic magnetic field configurations. We find that large values of $\beta \! \sim \! 200$ are required for typical toroidal fluxes of $\lambda^2 \! \sim \! 1-10$. This entails internal temperature, pressure, and flux perturbations that are $\sim$200 times larger than a naive estimate of $\sim \! B^2/ (G M^2/R^4)$.
%where $B_{\rm max}$ is the peak magnetic field within the star. Our solutions typically have $B_{\rm max} \! \sim \! 5 B_{\rm surf}$, 
%Hence the flux perturbations are $\delta L/L \! \sim \! 500 B_{\rm surf}^2/ (G M^2/R^4)$, larger than a naive estimate might predict.
Nonetheless, even for the strongest observed fields of $B \sim 10 \, {\rm kG}$ in main sequence stars, the bolometric luminosity variation is $\delta L/L \lesssim 10^{-6}$ (\autoref{fig:magstrucvis}) and is not detectable even with high-quality space-based photometry.

Applying von Zeipel's law near the surface of the star, one might expect magnetic fields to produce flux perturbations of order
\begin{equation}
    \frac{\delta L}{L} \sim \frac{f_{\rm mag}}{f_{\rm grav}}  \sim \frac{B_{\rm surf}^2}{4 \pi \rho R g} \, .
\end{equation}
This translates to 
\begin{equation}
    \label{eq:dlf}
    \frac{\delta L}{L} \sim 0.004 \bigg(\frac{B_{\rm surf}}{1 \, {\rm kG}}\bigg)^{\! \! 2} \! \! \bigg( \frac{\rho}{10^{-8} {\rm g/cm}^3} \bigg)^{\!\! -1} \! \!  \bigg(\frac{M}{M_\odot}\bigg)^{-1} \!\! \bigg(\frac{R}{R_\odot}\bigg) \ \, .
\end{equation}
This is several orders of magnitude larger than our calculations, clearly ruling out this expectation. Even though magnetic forces can be comparable to gravity near the star's surface, this applies only in a very thin layer near the photosphere. The deep interior (where the outgoing thermal flux is determined) has much higher density and is only weakly distorted, leading to a flux perturbation on the order of equation \ref{eq:dlsurf}, which is the von Zeipel expectation applied to the deep interior. The even more naive estimate 
\begin{equation}
    \label{eq:dlp}
    \frac{\delta L}{L} \sim \frac{P_{\rm mag}}{P_{\rm gas}} \sim 4  \bigg(\frac{B_{\rm surf}}{1 \, {\rm kG}}\bigg)^2 \bigg( \frac{P_{\rm gas}}{10^{4} {\rm erg \, cm}^{-3}} \bigg)^{-1} \ \, 
\end{equation}
can be ruled out for the same reasons.

It would be useful to relate the observed flux perturbation or surface magnetic field strength to a star's internal magnetic field strength. This will be difficult to accomplish from flux perturbations until a better understanding of their cause is established. Our results suggest that central magnetic field strengths can be anywhere from $\sim$3-50 larger than surface field strengths (see \autoref{fig:magstrucB}) which is in qualitative agreement with numerical simulations by \cite{Braithwaite2008} (see Figure 8 of that work), with higher central field strengths for higher magnetic forces or helicity. The force will be difficult to observationally quantify but the helicity could potentially be determined if the star's surface toroidal field can be measured \citep[e.g., Figure 6 in][]{Kochukhovetal2011}.

The effects of magnetic fields are very different for systems not in thermal equilibrium. In stars with transient magnetic activity (e.g., spots in magnetically active stars), magnetic spot life times can be much shorter than the star's local thermal time, depending on the depth of the spots. Magnetic fields can also temporarily disrupt convective energy transport, which also occurs on a thermal time. This is why the Sun's spots can appear dark: they are not in thermal equilibrium with underlying layers. In our work, we predict that magnetic poles of radiative stars can be either hot or cool, although we have argued they are more likely to be hot for realistic magnetic field configurations. This agrees with the sign predicted by \cite{Cantiello:11}, who examined spots in hydrostatic equilibrium but not thermal equilibrium. However, thermal diffusion could drastically reduce the flux perturbation below their estimate (essentially equation \ref{eq:dlp}), for spots that live longer than the local thermal time.

\subsection{Limitations}

For the most part, our methods are general and are applicable to nearly any type of radiative star, such as massive stars or white dwarfs. However, there are a few modifications that need to be made depending on the circumstances.

In this work, we did not compute the physical displacement vector $\bxi$, which does not appear anywhere in our set of equations. The reason for this is that the final state of the system (i.e., the perturbed pressure, temperature, etc.) and its final energy is independent of the displacements needed to reach that configuration. There are infinite combinations of $\xi_r$ and $\xi_\perp$ that satisfy the continuity equation 
\begin{equation}
    \delta \rho + \bnab \cdot \big( \rho \bxi \big) = 0 \, .
\end{equation}
However, we assumed uniform composition in our equation of state (equation \ref{eq:eos}), a good approximation for young main sequence stars. In stars with composition differences, the equation of state will contain an extra $\chi_\mu (\delta \mu/\mu)$ term, where $\mu$ is the mean molecular weight. If composition does not diffuse, we have $\delta \mu \sim -\xi_r d \mu/dr$, and hence the perturbed state will depend explicitly on the displacement vector. The equilibrium configuration is then presumably given by the displacement which minimizes the total energy of the perturbed system. This should be accounted for when considering stars with composition gradients (e.g., evolved stars and white dwarfs). 

Another issue we have neglected is anisotropic conduction induced by magnetic fields. In main sequence stars, this effect is only important in the surface layers where the electron mean-free path increases and becomes comparable to the Larmor radius. However, it may be important in the deep interiors of white dwarfs where electrons conduct most of the heat \citep{potekhin:99,potekhin:01,chang:10} and have fairly long mean free paths due to the high electron degeneracy. We hope to examine this effect in future work. 

We have also neglected any magnetically induced changes to opacity. These could be important near a star's surface because magnetic fields split the energy levels of atomic transitions, changing the opacity from bound-bound, bound-free absorption, and free-free absorption (e.g., \citealt{jordan:92}). We suspect that this will not alter our conclusions regarding the perturbation to the bolometric surface flux, because effects limited to the surface layers cannot change the emerging flux from below. This can be seen from equation \ref{eq:dldr}, because the second term is of order unity near the surface, and will only change the emerging flux by an amount $\sim \Delta r/r$, where $\Delta r$ is the width over which near-surface effects are important. 

What is more likely is that the star's emergent spectrum is altered by magnetic changes to opacity, or by composition differences between the magnetic pole and equator. Even with no perturbation to the bolometric flux, a changing spectrum could create large differences in, e.g., g-band or r-band fluxes as the magnetic pole rotates in and out of view. As an example, Caiazzo in prep. finds large changes in the composition and spectrum as a function of rotational phase in the magnetic WD ZTF J203349.8+322901.1 (``Janus"), even though there is no clear variation in the bolometric flux. Similar photometric variations (typically $\sim$1-3\% in amplitude) are observed in chemically peculiar magnetic A type stars \citep{hummerich:18}. Compositional inhomogeneities could naturally arise due to the perturbed gas pressure in the near-surface layers, which will alter atomic diffusion processes. This process should be studied in more detail.

\section{Conclusions}

We have computed the effects of strong magnetic fields on the structures of radiative main sequence stars. Our focus is the perturbed surface temperature and radiative flux induced by the magnetic field, which can produce photometric modulation as the star rotates. Unlike most prior work, we have computed structures in both hydrostatic and thermal equilibrium, which applies to stars with long-lived fossil fields, such as magnetic Ap stars. Our models have simple dipolar angular structure and include toroidal fields with associated magnetic helicity $\lambda$.

We find that magnetic fields at observed field strengths of $\sim \! 1 \, {\rm kG}$ produce negligible bolometric flux perturbations, $\delta L/L \lesssim 10^{-6}$. Even though such fields are large enough to produce significant perturbations to the photospheric gas pressure and hydrostatic force balance, the radiative flux is determined by deeper layers of the star where magnetic forces are negligible. The perturbed surface flux is comparable to the von Zeipel theorem estimate $\delta L/L \sim f_{\rm mag}/(\rho g)$ only when evaluated in the deep interior. Depending on their helicity and magnetic force, internal magnetic fields are typically a factor of $\sim$10 larger than surface magnetic fields.

The size of the magnetic perturbation depends on the strength of the magnetic forces, parameterized by $\beta$, which is zero for a force-free field. Relatively large values of $\beta \! \sim \! 200$ are needed for significant modification of the magnetic field profile relative to a force-free configuration. We have argued that these values of $\beta$ are most likely to occur in real stars such that the magnetic profile matches onto boundary conditions minimizing electric current near the surface. This leads to photometric modulations that are a few hundred times larger than a naive estimate of $\delta L/L \! \sim \! B_{\rm surf}^2 R^4/(G M^2)$, but still too small to be observed. Photometric modulations observed in magnetic stars likely arise from changes in the emergent spectrum rather than changes in the bolometric flux. Although we have focused on a young $3\, M_\odot$ model in this work, the same method can be applied to other types of predominantly radiative stars, such as moderately evolved massive stars or white dwarfs.

Our work can be improved in several ways. First, realistic magnetic configurations likely have toroidal fields confined to closed poloidal surfaces within the star such that current does not flow into the atmosphere and dissipate the magnetic field. Incorporating this condition will require more complicated magnetic topologies and non-separable solutions to the hydrostatic balance and radiative diffusion equations. The impacts  of rotation, where the rotation and magnetic axis are often misaligned, and of the associated centrifugal forces have also been neglected in this work \citep{Monaghan1973,Galea1985}. Our models do not account for composition gradients within a star, which may significantly affect the magnetic perturbations in evolved stars and white dwarfs. Finally, magnetic changes to opacity and anisotropic conduction should be included in future models in order to better interpret observable photometric and spectroscopic variations of magnetic stars.

\section*{Acknowledgments}

JF is thankful for support through an Innovator Grant from The Rose Hills Foundation. S.M. acknowledges support from CNES SOHO, PLATO, and LISA grants at CEA/IRFU/DAp.

\section*{Data Availability}

The relaxation code to compute magnetic perturbations is available upon request.

\bibliography{CoreRotBib,library}

\begin{thebibliography}{}
\makeatletter
\relax
\def\mn@urlcharsother{\let\do\@makeother \do\$\do\&\do\#\do\^\do\_\do\%\do\~}
\def\mn@doi{\begingroup\mn@urlcharsother \@ifnextchar [ {\mn@doi@}
  {\mn@doi@[]}}
\def\mn@doi@[#1]#2{\def\@tempa{#1}\ifx\@tempa\@empty \href
  {http://dx.doi.org/#2} {doi:#2}\else \href {http://dx.doi.org/#2} {#1}\fi
  \endgroup}
\def\mn@eprint#1#2{\mn@eprint@#1:#2::\@nil}
\def\mn@eprint@arXiv#1{\href {http://arxiv.org/abs/#1} {{\tt arXiv:#1}}}
\def\mn@eprint@dblp#1{\href {http://dblp.uni-trier.de/rec/bibtex/#1.xml}
  {dblp:#1}}
\def\mn@eprint@#1:#2:#3:#4\@nil{\def\@tempa {#1}\def\@tempb {#2}\def\@tempc
  {#3}\ifx \@tempc \@empty \let \@tempc \@tempb \let \@tempb \@tempa \fi \ifx
  \@tempb \@empty \def\@tempb {arXiv}\fi \@ifundefined
  {mn@eprint@\@tempb}{\@tempb:\@tempc}{\expandafter \expandafter \csname
  mn@eprint@\@tempb\endcsname \expandafter{\@tempc}}}

\bibitem[\protect\citeauthoryear{{Akg{\"u}n}, {Reisenegger}, {Mastrano}  \&
  {Marchant}}{{Akg{\"u}n} et~al.}{2013}]{akgun:13}
{Akg{\"u}n} T.,  {Reisenegger} A.,  {Mastrano} A.,   {Marchant} P.,  2013,
  \mn@doi [\mnras] {10.1093/mnras/stt913}, \href
  {https://ui.adsabs.harvard.edu/abs/2013MNRAS.433.2445A} {433, 2445}

\bibitem[\protect\citeauthoryear{{Becerra}, {Reisenegger}, {Valdivia}  \&
  {Gusakov}}{{Becerra} et~al.}{2022a}]{becerra:22}
{Becerra} L.,  {Reisenegger} A.,  {Valdivia} J.~A.,   {Gusakov} M.~E.,  2022a,
  \mn@doi [\mnras] {10.1093/mnras/stac102}, \href
  {https://ui.adsabs.harvard.edu/abs/2022MNRAS.511..732B} {511, 732}

\bibitem[\protect\citeauthoryear{{Becerra}, {Reisenegger}, {Valdivia}  \&
  {Gusakov}}{{Becerra} et~al.}{2022b}]{becerra:22b}
{Becerra} L.,  {Reisenegger} A.,  {Valdivia} J.~A.,   {Gusakov} M.,  2022b,
  \mn@doi [\mnras] {10.1093/mnras/stac2704}, \href
  {https://ui.adsabs.harvard.edu/abs/2022MNRAS.517..560B} {517, 560}

\bibitem[\protect\citeauthoryear{Braithwaite}{Braithwaite}{2008}]{Braithwaite2008}
Braithwaite J.,  2008, \mn@doi [Mon. Not. R. Astron. Soc.]
  {10.1111/j.1365-2966.2008.13218.x}, 386, 1947

\bibitem[\protect\citeauthoryear{{Braithwaite}}{{Braithwaite}}{2009}]{braithwaite:09}
{Braithwaite} J.,  2009, \mn@doi [\mnras] {10.1111/j.1365-2966.2008.14034.x},
  \href {http://adsabs.harvard.edu/abs/2009MNRAS.397..763B} {397, 763}

\bibitem[\protect\citeauthoryear{Braithwaite \& Nordlund}{Braithwaite \&
  Nordlund}{2006}]{Braithwaite_2006}
Braithwaite J.,  Nordlund {\AA}.,  2006, \mn@doi [Astronomy {\&} Astrophysics]
  {10.1051/0004-6361:20041980}, 450, 1077

\bibitem[\protect\citeauthoryear{Braithwaite \& Spruit}{Braithwaite \&
  Spruit}{2004}]{Braithwaite2004}
Braithwaite J.,  Spruit H.~C.,  2004, \mn@doi [Nature] {10.1038/nature02934},
  431, 819

\bibitem[\protect\citeauthoryear{{Braithwaite} \& {Spruit}}{{Braithwaite} \&
  {Spruit}}{2017}]{braithwaite:17}
{Braithwaite} J.,  {Spruit} H.~C.,  2017, \mn@doi [Royal Society Open Science]
  {10.1098/rsos.160271}, \href
  {https://ui.adsabs.harvard.edu/abs/2017RSOS....460271B} {4, 160271}

\bibitem[\protect\citeauthoryear{{Broderick} \& {Narayan}}{{Broderick} \&
  {Narayan}}{2008}]{broderick:08}
{Broderick} A.~E.,  {Narayan} R.,  2008, \mn@doi [\mnras]
  {10.1111/j.1365-2966.2007.12634.x}, \href
  {https://ui.adsabs.harvard.edu/abs/2008MNRAS.383..943B} {383, 943}

\bibitem[\protect\citeauthoryear{{Bugnet} et~al.,}{{Bugnet}
  et~al.}{2021}]{Bugnet:2021}
{Bugnet} L.,  et~al., 2021, \mn@doi [\aap] {10.1051/0004-6361/202039159}, \href
  {https://ui.adsabs.harvard.edu/abs/2021A&A...650A..53B} {650, A53}

\bibitem[\protect\citeauthoryear{{Busse}}{{Busse}}{1981}]{Busse:1981}
{Busse} F.~H.,  1981, \mn@doi [Geophysical and Astrophysical Fluid Dynamics]
  {10.1080/03091928108243683}, \href
  {https://ui.adsabs.harvard.edu/abs/1981GApFD..17..215B} {17, 215}

\bibitem[\protect\citeauthoryear{{Cantiello} \& {Braithwaite}}{{Cantiello} \&
  {Braithwaite}}{2011}]{Cantiello:11}
{Cantiello} M.,  {Braithwaite} J.,  2011, \mn@doi [\aap]
  {10.1051/0004-6361/201117512}, \href
  {https://ui.adsabs.harvard.edu/abs/2011A&A...534A.140C} {534, A140}

\bibitem[\protect\citeauthoryear{{Chandrasekhar}}{{Chandrasekhar}}{1956}]{chandrasekhar:56}
{Chandrasekhar} S.,  1956, \mn@doi [\apj] {10.1086/146217}, \href
  {https://ui.adsabs.harvard.edu/abs/1956ApJ...124..232C} {124, 232}

\bibitem[\protect\citeauthoryear{{Chandrasekhar} \&
  {Prendergast}}{{Chandrasekhar} \& {Prendergast}}{1956}]{chandrasekhar:56b}
{Chandrasekhar} S.,  {Prendergast} K.~H.,  1956, \mn@doi [Proceedings of the
  National Academy of Science] {10.1073/pnas.42.1.5}, \href
  {https://ui.adsabs.harvard.edu/abs/1956PNAS...42....5C} {42, 5}

\bibitem[\protect\citeauthoryear{{Chang} \& {Quataert}}{{Chang} \&
  {Quataert}}{2010}]{chang:10}
{Chang} P.,  {Quataert} E.,  2010, \mn@doi [\mnras]
  {10.1111/j.1365-2966.2009.15756.x}, \href
  {https://ui.adsabs.harvard.edu/abs/2010MNRAS.403..246C} {403, 246}

\bibitem[\protect\citeauthoryear{{Davies}}{{Davies}}{1968}]{davies:68}
{Davies} G.~F.,  1968, \mn@doi [Australian Journal of Physics]
  {10.1071/PH680293}, \href
  {https://ui.adsabs.harvard.edu/abs/1968AuJPh..21..293D} {21, 293}

\bibitem[\protect\citeauthoryear{{Decressin}, {Mathis}, {Palacios}, {Siess},
  {Talon}, {Charbonnel}  \& {Zahn}}{{Decressin} et~al.}{2009}]{decressin:2009}
{Decressin} T.,  {Mathis} S.,  {Palacios} A.,  {Siess} L.,  {Talon} S.,
  {Charbonnel} C.,   {Zahn} J.~P.,  2009, \mn@doi [\aap]
  {10.1051/0004-6361:200810665}, \href
  {https://ui.adsabs.harvard.edu/abs/2009A&A...495..271D} {495, 271}

\bibitem[\protect\citeauthoryear{{Duez} \& {Mathis}}{{Duez} \&
  {Mathis}}{2010}]{duez:10}
{Duez} V.,  {Mathis} S.,  2010, \mn@doi [\aap] {10.1051/0004-6361/200913496},
  \href {https://ui.adsabs.harvard.edu/abs/2010A&A...517A..58D} {517, A58}

\bibitem[\protect\citeauthoryear{{Duez}, {Mathis}  \&
  {Turck-Chi{\`e}ze}}{{Duez} et~al.}{2010a}]{duez:10c}
{Duez} V.,  {Mathis} S.,   {Turck-Chi{\`e}ze} S.,  2010a, \mn@doi [\mnras]
  {10.1111/j.1365-2966.2009.15955.x}, \href
  {https://ui.adsabs.harvard.edu/abs/2010MNRAS.402..271D} {402, 271}

\bibitem[\protect\citeauthoryear{{Duez}, {Braithwaite}  \& {Mathis}}{{Duez}
  et~al.}{2010b}]{duez:10b}
{Duez} V.,  {Braithwaite} J.,   {Mathis} S.,  2010b, \mn@doi [\apjl]
  {10.1088/2041-8205/724/1/L34}, \href
  {https://ui.adsabs.harvard.edu/abs/2010ApJ...724L..34D} {724, L34}

\bibitem[\protect\citeauthoryear{{Eddington}}{{Eddington}}{1929}]{eddington:29}
{Eddington} A.~S.,  1929, \mn@doi [\mnras] {10.1093/mnras/90.1.54}, \href
  {https://ui.adsabs.harvard.edu/abs/1929MNRAS..90...54E} {90, 54}

\bibitem[\protect\citeauthoryear{{Fuller}, {Cantiello}, {Stello}, {Garcia}  \&
  {Bildsten}}{{Fuller} et~al.}{2015}]{Fulleretal2015}
{Fuller} J.,  {Cantiello} M.,  {Stello} D.,  {Garcia} R.~A.,   {Bildsten} L.,
  2015, \mn@doi [Science] {10.1126/science.aac6933}, \href
  {https://ui.adsabs.harvard.edu/abs/2015Sci...350..423F} {350, 423}

\bibitem[\protect\citeauthoryear{{Galea} \& {Wood}}{{Galea} \&
  {Wood}}{1985}]{Galea1985}
{Galea} E.~R.,  {Wood} W.~P.,  1985, \mn@doi [\mnras]
  {10.1093/mnras/217.3.633}, \href
  {https://ui.adsabs.harvard.edu/abs/1985MNRAS.217..633G} {217, 633}

\bibitem[\protect\citeauthoryear{Garc{\'{i}}a et~al.,}{Garc{\'{i}}a
  et~al.}{2014}]{Garcia2014}
Garc{\'{i}}a R.~A.,  et~al., 2014, \mn@doi [Astron. Astrophys.]
  {10.1051/0004-6361/201322823}, 563, A84

\bibitem[\protect\citeauthoryear{{H{\"u}mmerich} et~al.,}{{H{\"u}mmerich}
  et~al.}{2018}]{hummerich:18}
{H{\"u}mmerich} S.,  et~al., 2018, \mn@doi [\aap]
  {10.1051/0004-6361/201832938}, \href
  {https://ui.adsabs.harvard.edu/abs/2018A&A...619A..98H} {619, A98}

\bibitem[\protect\citeauthoryear{{Jordan}}{{Jordan}}{1992}]{jordan:92}
{Jordan} S.,  1992, \aap, \href
  {https://ui.adsabs.harvard.edu/abs/1992A&A...265..570J} {265, 570}

\bibitem[\protect\citeauthoryear{{Kaufman}, {Lecoanet}, {Anders}, {Brown},
  {Vasil}, {Oishi}  \& {Burns}}{{Kaufman} et~al.}{2022}]{Kaufman:2022}
{Kaufman} E.,  {Lecoanet} D.,  {Anders} E.~H.,  {Brown} B.~P.,  {Vasil} G.~M.,
  {Oishi} J.~S.,   {Burns} K.~J.,  2022, \mn@doi [\mnras]
  {10.1093/mnras/stac2707}, \href
  {https://ui.adsabs.harvard.edu/abs/2022MNRAS.517.3332K} {517, 3332}

\bibitem[\protect\citeauthoryear{{Kochukhov}, {Lundin}, {Romanyuk}  \&
  {Kudryavtsev}}{{Kochukhov} et~al.}{2011}]{Kochukhovetal2011}
{Kochukhov} O.,  {Lundin} A.,  {Romanyuk} I.,   {Kudryavtsev} D.,  2011,
  \mn@doi [\apj] {10.1088/0004-637X/726/1/24}, \href
  {https://ui.adsabs.harvard.edu/abs/2011ApJ...726...24K} {726, 24}

\bibitem[\protect\citeauthoryear{{Lander} \& {Jones}}{{Lander} \&
  {Jones}}{2012}]{lander:12}
{Lander} S.~K.,  {Jones} D.~I.,  2012, \mn@doi [\mnras]
  {10.1111/j.1365-2966.2012.21213.x}, \href
  {https://ui.adsabs.harvard.edu/abs/2012MNRAS.424..482L} {424, 482}

\bibitem[\protect\citeauthoryear{{Landstreet}}{{Landstreet}}{1987}]{landstreet:87}
{Landstreet} J.~D.,  1987, \mn@doi [\mnras] {10.1093/mnras/225.2.437}, \href
  {https://ui.adsabs.harvard.edu/abs/1987MNRAS.225..437L} {225, 437}

\bibitem[\protect\citeauthoryear{{Landstreet} \& {Mathys}}{{Landstreet} \&
  {Mathys}}{2000}]{LandstreetMathys2000}
{Landstreet} J.~D.,  {Mathys} G.,  2000, \aap, \href
  {https://ui.adsabs.harvard.edu/abs/2000A&A...359..213L} {359, 213}

\bibitem[\protect\citeauthoryear{{Lecoanet}, {Vasil}, {Fuller}, {Cantiello}  \&
  {Burns}}{{Lecoanet} et~al.}{2017}]{Lecoanetetal2017}
{Lecoanet} D.,  {Vasil} G.~M.,  {Fuller} J.,  {Cantiello} M.,   {Burns} K.~J.,
  2017, \mn@doi [\mnras] {10.1093/mnras/stw3273}, \href
  {https://ui.adsabs.harvard.edu/abs/2017MNRAS.466.2181L} {466, 2181}

\bibitem[\protect\citeauthoryear{{Li}, {Ventura}, {Basu}, {Sofia}  \&
  {Demarque}}{{Li} et~al.}{2006}]{li:06}
{Li} L.~H.,  {Ventura} P.,  {Basu} S.,  {Sofia} S.,   {Demarque} P.,  2006,
  \mn@doi [\apjs] {10.1086/502800}, \href
  {https://ui.adsabs.harvard.edu/abs/2006ApJS..164..215L} {164, 215}

\bibitem[\protect\citeauthoryear{{Li}, {Deheuvels}, {Ballot}  \&
  {Ligni{\`e}res}}{{Li} et~al.}{2022}]{Li:2022}
{Li} G.,  {Deheuvels} S.,  {Ballot} J.,   {Ligni{\`e}res} F.,  2022, \mn@doi
  [\nat] {10.1038/s41586-022-05176-0}, \href
  {https://ui.adsabs.harvard.edu/abs/2022Natur.610...43L} {610, 43}

\bibitem[\protect\citeauthoryear{{Loi}}{{Loi}}{2021}]{Loi:2021}
{Loi} S.~T.,  2021, \mn@doi [\mnras] {10.1093/mnras/stab991}, \href
  {https://ui.adsabs.harvard.edu/abs/2021MNRAS.504.3711L} {504, 3711}

\bibitem[\protect\citeauthoryear{{Lyutikov}}{{Lyutikov}}{2010}]{lyutikov:10}
{Lyutikov} M.,  2010, \mn@doi [\mnras] {10.1111/j.1365-2966.2009.15876.x},
  \href {https://ui.adsabs.harvard.edu/abs/2010MNRAS.402..345L} {402, 345}

\bibitem[\protect\citeauthoryear{{Maeder}}{{Maeder}}{1999}]{maeder:99}
{Maeder} A.,  1999, \aap, \href
  {https://ui.adsabs.harvard.edu/abs/1999A&A...347..185M} {347, 185}

\bibitem[\protect\citeauthoryear{{Markey} \& {Tayler}}{{Markey} \&
  {Tayler}}{1973}]{markeytayler:73}
{Markey} P.,  {Tayler} R.~J.,  1973, \mn@doi [\mnras] {10.1093/mnras/163.1.77},
  \href {https://ui.adsabs.harvard.edu/abs/1973MNRAS.163...77M} {163, 77}

\bibitem[\protect\citeauthoryear{{Mathis}}{{Mathis}}{2013}]{mathis:2013}
{Mathis} S.,  2013, in {Goupil} M.,  {Belkacem} K.,  {Neiner} C.,
  {Ligni{\`e}res} F.,   {Green} J.~J.,  eds, , Vol.~865, Lecture Notes in
  Physics, Berlin Springer Verlag.
p.~23, \mn@doi{10.1007/978-3-642-33380-4_2}

\bibitem[\protect\citeauthoryear{{Mathis}, {Bugnet}, {Prat}, {Augustson},
  {Mathur}  \& {Garcia}}{{Mathis} et~al.}{2021}]{Mathis:2021}
{Mathis} S.,  {Bugnet} L.,  {Prat} V.,  {Augustson} K.,  {Mathur} S.,
  {Garcia} R.~A.,  2021, \mn@doi [\aap] {10.1051/0004-6361/202039180}, \href
  {https://ui.adsabs.harvard.edu/abs/2021A&A...647A.122M} {647, A122}

\bibitem[\protect\citeauthoryear{{Mestel} \& {Moss}}{{Mestel} \&
  {Moss}}{1977}]{mestel:77}
{Mestel} L.,  {Moss} D.~L.,  1977, \mn@doi [\mnras] {10.1093/mnras/178.1.27},
  \href {https://ui.adsabs.harvard.edu/abs/1977MNRAS.178...27M} {178, 27}

\bibitem[\protect\citeauthoryear{{Milsom} \& {Wright}}{{Milsom} \&
  {Wright}}{1976}]{milsom:76}
{Milsom} F.,  {Wright} G.~A.~E.,  1976, \mn@doi [\mnras]
  {10.1093/mnras/174.2.307}, \href
  {https://ui.adsabs.harvard.edu/abs/1976MNRAS.174..307M} {174, 307}

\bibitem[\protect\citeauthoryear{{Monaghan}}{{Monaghan}}{1966}]{monaghan:66}
{Monaghan} F.~F.,  1966, \mn@doi [\mnras] {10.1093/mnras/132.1.1}, \href
  {https://ui.adsabs.harvard.edu/abs/1966MNRAS.132....1M} {132, 1}

\bibitem[\protect\citeauthoryear{{Monaghan}}{{Monaghan}}{1973}]{Monaghan1973}
{Monaghan} J.~J.,  1973, \mn@doi [\mnras] {10.1093/mnras/163.4.423}, \href
  {https://ui.adsabs.harvard.edu/abs/1973MNRAS.163..423M} {163, 423}

\bibitem[\protect\citeauthoryear{{Morel} et~al.,}{{Morel}
  et~al.}{2014}]{Morel:2014}
{Morel} T.,  et~al., 2014, The Messenger, \href
  {https://ui.adsabs.harvard.edu/abs/2014Msngr.157...27M} {157, 27}

\bibitem[\protect\citeauthoryear{{Moss}}{{Moss}}{1973}]{moss:1973}
{Moss} D.~L.,  1973, \mn@doi [\mnras] {10.1093/mnras/164.1.33}, \href
  {https://ui.adsabs.harvard.edu/abs/1973MNRAS.164...33M} {164, 33}

\bibitem[\protect\citeauthoryear{{Moss}}{{Moss}}{1979}]{moss:79}
{Moss} D.,  1979, \mn@doi [\mnras] {10.1093/mnras/187.3.601}, \href
  {https://ui.adsabs.harvard.edu/abs/1979MNRAS.187..601M} {187, 601}

\bibitem[\protect\citeauthoryear{{Oksala}, {Silvester}, {Kochukhov}, {Neiner},
  {Wade}  \& {MiMeS Collaboration}}{{Oksala} et~al.}{2018}]{oksala:18}
{Oksala} M.~E.,  {Silvester} J.,  {Kochukhov} O.,  {Neiner} C.,  {Wade} G.~A.,
   {MiMeS Collaboration} 2018, \mn@doi [\mnras] {10.1093/mnras/stx2487}, \href
  {https://ui.adsabs.harvard.edu/abs/2018MNRAS.473.3367O} {473, 3367}

\bibitem[\protect\citeauthoryear{{Ostriker} \& {Hartwick}}{{Ostriker} \&
  {Hartwick}}{1968}]{ostriker:68}
{Ostriker} J.~P.,  {Hartwick} F.~D.~A.,  1968, \mn@doi [\apj] {10.1086/149706},
  \href {https://ui.adsabs.harvard.edu/abs/1968ApJ...153..797O} {153, 797}

\bibitem[\protect\citeauthoryear{{Paxton}, {Bildsten}, {Dotter}, {Herwig},
  {Lesaffre}  \& {Timmes}}{{Paxton} et~al.}{2011}]{paxton:11}
{Paxton} B.,  {Bildsten} L.,  {Dotter} A.,  {Herwig} F.,  {Lesaffre} P.,
  {Timmes} F.,  2011, \mn@doi [\apjs] {10.1088/0067-0049/192/1/3}, \href
  {http://adsabs.harvard.edu/abs/2011ApJS..192....3P} {192, 3}

\bibitem[\protect\citeauthoryear{{Potekhin}}{{Potekhin}}{1999}]{potekhin:99}
{Potekhin} A.~Y.,  1999, \aap, \href
  {https://ui.adsabs.harvard.edu/abs/1999A&A...351..787P} {351, 787}

\bibitem[\protect\citeauthoryear{{Potekhin} \& {Yakovlev}}{{Potekhin} \&
  {Yakovlev}}{2001}]{potekhin:01}
{Potekhin} A.~Y.,  {Yakovlev} D.~G.,  2001, \mn@doi [\aap]
  {10.1051/0004-6361:20010698}, \href
  {https://ui.adsabs.harvard.edu/abs/2001A&A...374..213P} {374, 213}

\bibitem[\protect\citeauthoryear{Press, Teukolsky, Vetterling  \&
  Flannery}{Press et~al.}{2007}]{press:07}
Press W.~H.,  Teukolsky S.~A.,  Vetterling W.~T.,   Flannery B.~P.,  2007,
  Numerical Recipes 3rd Edition: The Art of Scientific Computing, 3 edn.
Cambridge University Press, USA

\bibitem[\protect\citeauthoryear{{Raadu}}{{Raadu}}{1971}]{raadu:71}
{Raadu} M.~A.,  1971, \mn@doi [\apss] {10.1007/BF00653334}, \href
  {https://ui.adsabs.harvard.edu/abs/1971Ap&SS..14..464R} {14, 464}

\bibitem[\protect\citeauthoryear{{Reisenegger}}{{Reisenegger}}{2009}]{Reisenegger:2009}
{Reisenegger} A.,  2009, \mn@doi [\aap] {10.1051/0004-6361/200810895}, \href
  {https://ui.adsabs.harvard.edu/abs/2009A&A...499..557R} {499, 557}

\bibitem[\protect\citeauthoryear{Rieutord}{Rieutord}{2006}]{Rieutord2006}
Rieutord M.,  2006, \mn@doi [EAS Publ. Ser.] {10.1051/eas:2006117}, 21, 275

\bibitem[\protect\citeauthoryear{{Shultz} et~al.,}{{Shultz}
  et~al.}{2019}]{Shultz:2019}
{Shultz} M.~E.,  et~al., 2019, \mn@doi [\mnras] {10.1093/mnras/stz2551}, \href
  {https://ui.adsabs.harvard.edu/abs/2019MNRAS.490..274S} {490, 274}

\bibitem[\protect\citeauthoryear{{Shulyak} et~al.,}{{Shulyak}
  et~al.}{2007}]{Shulyak:2007}
{Shulyak} D.,  et~al., 2007, \mn@doi [\aap] {10.1051/0004-6361:20064998}, \href
  {https://ui.adsabs.harvard.edu/abs/2007A&A...464.1089S} {464, 1089}

\bibitem[\protect\citeauthoryear{{Shulyak}, {Kochukhov}, {Valyavin}, {Lee},
  {Galazutdinov}, {Kim}, {Han}  \& {Burlakova}}{{Shulyak}
  et~al.}{2010}]{Shulyak:2010}
{Shulyak} D.,  {Kochukhov} O.,  {Valyavin} G.,  {Lee} B.~C.,  {Galazutdinov}
  G.,  {Kim} K.~M.,  {Han} I.,   {Burlakova} T.,  2010, \mn@doi [\aap]
  {10.1051/0004-6361/200912615}, \href
  {https://ui.adsabs.harvard.edu/abs/2010A&A...509A..28S} {509, A28}

\bibitem[\protect\citeauthoryear{{Stello}, {Cantiello}, {Fuller}, {Huber},
  {Garc{\'{\i}}a}, {Bedding}, {Bildsten}  \& {Silva Aguirre}}{{Stello}
  et~al.}{2016}]{stello:16}
{Stello} D.,  {Cantiello} M.,  {Fuller} J.,  {Huber} D.,  {Garc{\'{\i}}a}
  R.~A.,  {Bedding} T.~R.,  {Bildsten} L.,   {Silva Aguirre} V.,  2016, \mn@doi
  [\nat] {10.1038/nature16171}, \href
  {http://adsabs.harvard.edu/abs/2016Natur.529..364S} {529, 364}

\bibitem[\protect\citeauthoryear{{Sweet}}{{Sweet}}{1950}]{sweet:50}
{Sweet} P.~A.,  1950, \mn@doi [\mnras] {10.1093/mnras/110.6.548}, \href
  {https://ui.adsabs.harvard.edu/abs/1950MNRAS.110..548S} {110, 548}

\bibitem[\protect\citeauthoryear{{Tayler}}{{Tayler}}{1973}]{tayler:73}
{Tayler} R.~J.,  1973, \mn@doi [\mnras] {10.1093/mnras/161.4.365}, \href
  {http://adsabs.harvard.edu/abs/1973MNRAS.161..365T} {161, 365}

\bibitem[\protect\citeauthoryear{{Tayler}}{{Tayler}}{1980}]{tayler:80}
{Tayler} R.~J.,  1980, \mn@doi [\mnras] {10.1093/mnras/191.1.151}, \href
  {https://ui.adsabs.harvard.edu/abs/1980MNRAS.191..151T} {191, 151}

\bibitem[\protect\citeauthoryear{{Taylor}}{{Taylor}}{1974}]{taylor:1974}
{Taylor} J.~B.,  1974, \mn@doi [\prl] {10.1103/PhysRevLett.33.1139}, \href
  {https://ui.adsabs.harvard.edu/abs/1974PhRvL..33.1139T} {33, 1139}

\bibitem[\protect\citeauthoryear{{Wade} et~al.,}{{Wade}
  et~al.}{2016}]{Wade:2016}
{Wade} G.~A.,  et~al., 2016, \mn@doi [\mnras] {10.1093/mnras/stv2568}, \href
  {https://ui.adsabs.harvard.edu/abs/2016MNRAS.456....2W} {456, 2}

\bibitem[\protect\citeauthoryear{{Wright}}{{Wright}}{1969}]{wright:69}
{Wright} G.~A.~E.,  1969, \mn@doi [\mnras] {10.1093/mnras/146.2.197}, \href
  {https://ui.adsabs.harvard.edu/abs/1969MNRAS.146..197W} {146, 197}

\bibitem[\protect\citeauthoryear{{von Zeipel}}{{von
  Zeipel}}{1924}]{vonZeipel:1924}
{von Zeipel} H.,  1924, \mn@doi [\mnras] {10.1093/mnras/84.9.665}, \href
  {https://ui.adsabs.harvard.edu/abs/1924MNRAS..84..665V} {84, 665}

\makeatother
\end{thebibliography}

%\appendix

%\section{Extra}

%Recall equation \ref{eq:hydror5}:
%\begin{equation}
%\label{eq:hydror6}
%\delta p \frac{\partial \Psi}{\partial r} = \bigg[\frac{\partial \delta p}{\partial r} + g \delta \rho \bigg] \Psi \, .
%\end{equation}

\end{document}